# Universal rules for visible-light absorption in hybrid perovskite materials

Masato Kato,[1] Takemasa Fujiseki,[1] Tetsuhiko Miyadera,[2] Takeshi Sugita,[2] Shohei Fujimoto,[1] Masato Tamakoshi,[1] Masayuki Chikamatsu,[2] and Hiroyuki Fujiwara[1,a)]

[1]*Department of Electrical, Electronic and Computer Engineering, Gifu University, 1-1 Yanagido, Gifu 501-1193, Japan*
[2]*Research Center for Photovoltaics, National Institute of Advanced Industrial Science and Technology (AIST), Central 5, 1-1-1 Higashi, Tsukuba, Ibaraki 305-8568, Japan*

## Abstract

A variety of organic-inorganic hybrid perovskites ($APbX_3$) consisting of mixed center cations [A = $CH_3NH_3^+$, $HC(NH_2)_2^+$, $Cs^+$] with different $PbX_3^-$ cages (X = I, Br, Cl) have been developed to realize high-efficiency solar cells. Nevertheless, clear understanding for the effects of A and X on the optical transition has been lacking. Here, we present universal rules that allow the unified interpretation of the optical absorption in various hybrid perovskites. In particular, we find that the influence of the A-site cation on the light absorption is rather significant and the absorption coefficient ($\alpha$) reduces to half when $CH_3NH_3^+$ is replaced with $HC(NH_2)_2^+$ in the $APbI_3$ system. Our density functional theory (DFT) calculations reproduce all of the fine absorption features observed in $HC(NH_2)_2PbI_3$ and $CH_3NH_3PbBr_3$, allowing the unique assignment of the interband transitions in the Brillouin zone. In contrast to general understanding that the A-site cation involves weakly in the optical process, our theoretical calculations reveal that the center cation plays a critical role in the interband transition and the absorption strength in the visible region is modified by the strong A-X interaction. Furthermore, our systematic analyses show that the variation of the absorption spectrum with X can be described simply by the well-known sum rule. The universal rules established in this study explain the large reduction of $\alpha$ in $HC(NH_2)_2PbI_3$ and predict $CsPbI_3$ as the highest $\alpha$ material.

## I. INTRODUCTION

Numerous organic-inorganic hybrid perovskites expressed by $APbX_3$ (A: organic cation; X: halogen atom) have been investigated[1-69] in an effort to develop highly stable solar cells with high efficiencies. Although quite extensive researches have been conducted on methylammonium lead iodide ($MAPbI_3$) perovskite,[1-44,68,69] now it becomes quite clear that the $MAPbI_3$ perovskite exhibits large instability for temperature,[3-8] light illumination,[6-9] and humidity.[6-11] In particular, $MAPbI_3$ shows a decomposition into $PbI_2$ at temperatures of ≥ 85 °C,[4,5] which could limit the practical application of this material severely.

The thermal stability can be improved greatly when the MA cation ($CH_3NH_3^+$) of $MAPbI_3$ is replaced with a formamidinium (FA) cation [$HC(NH_2)_2^+$] having a larger molecular weight. For $FAPbI_3$, no major structural change occurs upon thermal treatment up to 150 °C.[46] However, the $FAPbI_3$ perovskite has limited long-term stability and a cubic $FAPbI_3$ crystal ($\alpha$-$FAPbI_3$) shows a gradual phase transformation into a transparent $\delta$-$FAPbI_3$ phase having a one-dimensional crystal structure.[49,50] Such instability is caused by the larger size of $FA^+$ and, quite fortunately, $FAPbI_3$-based perovskites can be stabilized by including a small amount of $MA^+$ and $Cs^+$ having smaller ionic radii.[51-57] The Cs addition to $\alpha$-$FAPbI_3$ is also beneficial for suppressing degradation induced by humid air and light illumination.[53] A recent study further demonstrates the improved overall stability of $\alpha$-$FAPbI_3$ by the incorporation of Cs and Br atoms.[57] Accordingly, by the optimum combination of different A-site cations and X-site halogen atoms, an ideal hybrid perovskite compound with high stability could be realized. To date, very high conversion efficiencies exceeding 20% have been demonstrated in (FA, MA)Pb(I, Br)$_3$ and (FA, MA, Cs)Pb(I, Br)$_3$ solar cells.[52,57-59]

Nevertheless, despite the rapid progress for the solar cell fabrication, the optical process in the complex hybrid perovskite remains unclear. For the light absorption in $APbX_3$, there is a common belief that the A-site cation plays a minor role in the optical transition[12-19] and only the band gap ($E_g$) changes slightly according to the size of the A-site cation.[17,20] However, although many studies have been devoted to determine the dielectric functions ($\varepsilon = \varepsilon_1 - i\varepsilon_2$) of $MAPbI_3$ (Refs. 5, 23-27), $MAPbBr_3$ (Refs. 27, 63) and $MAPbCl_3$ (Ref. 27), only limited experimental results are available for the quantitative effect of the A-site cation on the light absorption.[53,60,61] So far, the optical properties of $FAPbI_3$ (Refs. 20, 65) and $CsPbI_3$ (Refs. 65-67) perovskites have been investigated by applying density functional theory (DFT), but the influences of the A-site cation and X-site halogen atom on the absorption properties remain ambiguous.

Due to the increasing complexity of hybrid perovskite compounds, there is a pressing need for quantitative interpretation of the light absorption in various hybrid perovskites.

In this article, we establish the universal rules that define the optical effects of A and X in $APbX_3$ perovskite materials (A = $MA^+$, $FA^+$, $Cs^+$; X = I, Br, Cl). In particular, we experimentally find a large variation of the absorption coefficient ($\alpha$) with the A-site cation for $APbI_3$. Our systematic DFT calculation reveals that $\alpha$ of the hybrid perovskites in the visible region is determined by the quite strong interaction of the X-atom valence electron with the A-site cation. Moreover, we present that the effect of X on the perovskite dielectric function can be expressed simply by the general sum rule.[70] Based on the guiding principles established in this study, $\alpha$ spectra of complex hybrid perovskites at room temperature can be predicted.

## II. EXPERIMENT

To determine the dielectric function of $\alpha$-$FAPbI_3$, we prepare $\alpha$-$FAPbI_3$ layers by a laser evaporation technique,[5] in which $PbI_2$ and FAI source materials are heated by a near-infrared laser with a wavelength of 808 nm. Our earlier study on $MAPbI_3$ confirms that the laser evaporation method is quite helpful for the preparation of smooth hybrid perovskite layers, which are quite important for reliable optical analyses.[5] In the laser evaporation process, the evaporation rates of the source materials are controlled by adjusting the laser power. The resulting deposition rate of $FAPbI_3$ layers is 2.3 nm/min. The laser evaporation process is conducted without substrate heating at a pressure of $4\times10^{-4}$ Pa.

We find that $\alpha$-$FAPbI_3$ and $\delta$-$FAPbI_3$ layers can be formed separately by choosing proper substrates. For the preparation of $\alpha$-$FAPbI_3$, we employ crystalline Si (*c*-Si) substrates covered with native oxides. For the deposition of $\alpha$-$FAPbI_3$, we particularly use uncleaned $SiO_2$/*c*-Si substrates, as ultrasonic cleaning of the substrates using organic solvents tends to enhance the $PbI_2$ formation. The spectroscopic ellipsometry (SE) analyses of the cleaned and uncleaned substrates show that the thickness of an organic overlayer on the uncleaned substrate is ~1 Å. On the other hand, $\delta$-$FAPbI_3$ can be formed by using substrates in which thin PCDTBT (poly[*N*-9''-hepta-decanyl-2,7-carbazole-alt-5,5-(4',7'-di-2-thienyl-2',1',3'-benzothiadiazole)]) layers (~5 nm) are formed on ZnO-coated *c*-Si substrates. The ZnO layer (50 nm) is provided to improve the adhesion of the PCDTBT layer on the substrate.

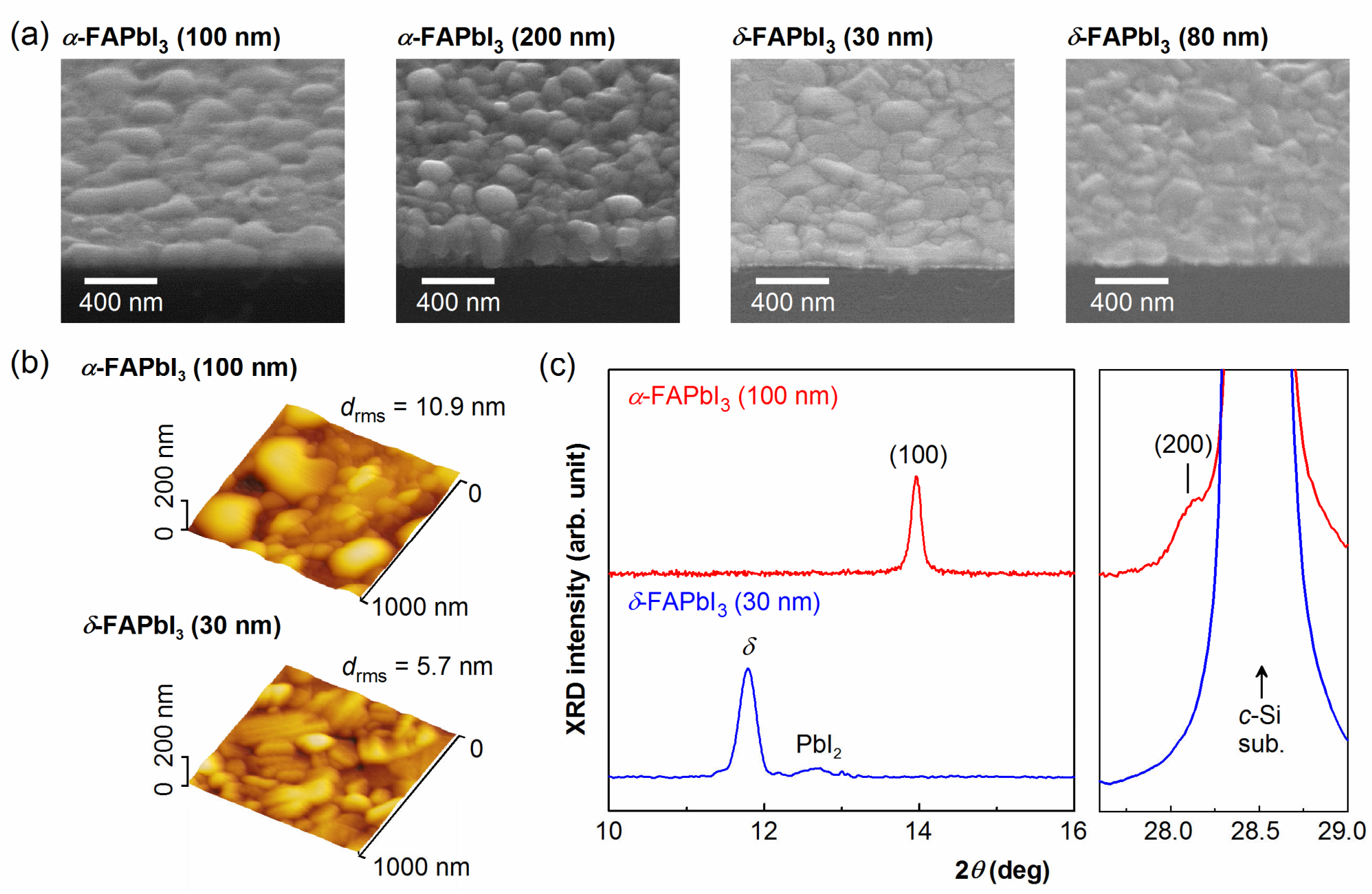

FIG. 1. (a) SEM images, (b) AFM images and (c) XRD spectra of the $\alpha$-$FAPbI_3$ and $\delta$-$FAPbI_3$ layers with different thicknesses, fabricated by the laser evaporation. The $\alpha$-$FAPbI_3$ and $\delta$-$FAPbI_3$ layers are prepared separately by using $SiO_2$(2 nm)/*c*-Si and PCDTBT(5 nm)/ZnO (50 nm)/*c*-Si substrates, respectively. In (b), the root-mean-square roughness ($d_{rms}$) obtained from each AFM image is also indicated.

Figure 1 shows (a) scanning electron microscope (SEM) images, (b) atomic force microscope (AFM) images and (c) x-ray diffraction (XRD) spectra of the $\alpha$-$FAPbI_3$ and $\delta$-$FAPbI_3$ layers prepared on the $SiO_2$/*c*-Si and PCDTBT/ZnO/c-Si substrates, respectively. The SEM images in Fig. 1(a) show significant surface roughening with increasing $\alpha$-$FAPbI_3$ layer thickness from 100 to 200 nm, and the preparation of smooth surface is more difficult for $\alpha$-$FAPbI_3$, compared with $MAPbI_3$. In fact, the root-mean-square roughness ($d_{rms}$) of the thin $\alpha$-$FAPbI_3$ layer (100 nm) estimated by AFM is relatively large [$d_{rms}$ = 10.9 nm in Fig. 1(b)] when compared with $d_{rms}$ = 4.6 nm observed for a laser-evaporated $MAPbI_3$ layer.[5] In the thick $\alpha$-$FAPbI_3$ layer (200 nm), $d_{rms}$ increases further to 12.6 nm.

The XRD spectrum obtained from the thin $\alpha$-$FAPbI_3$ layer [Fig. 1(c)] shows the sharp peaks attributed to (100) and (200) diffractions of the cubic $\alpha$ phase.[50] On the other hand, the $\delta$-$FAPbI_3$ layer with a thickness of 30 nm in Fig. 1(a) exhibits diffraction peaks at 11.8$^{o}$ and 12.7$^{o}$ due to the formation of $\delta$-$FAPbI_3$ (Refs. 3, 45, 48) and $PbI_2$ (Ref. 11), respectively [Fig. 1(c)]. The XRD result of Fig. 1(c) confirms that, in the $\alpha$-$FAPbI_3$ layer, the generation of the $\delta$-$FAPbI_3$ and $PbI_2$ secondary phases is negligible.

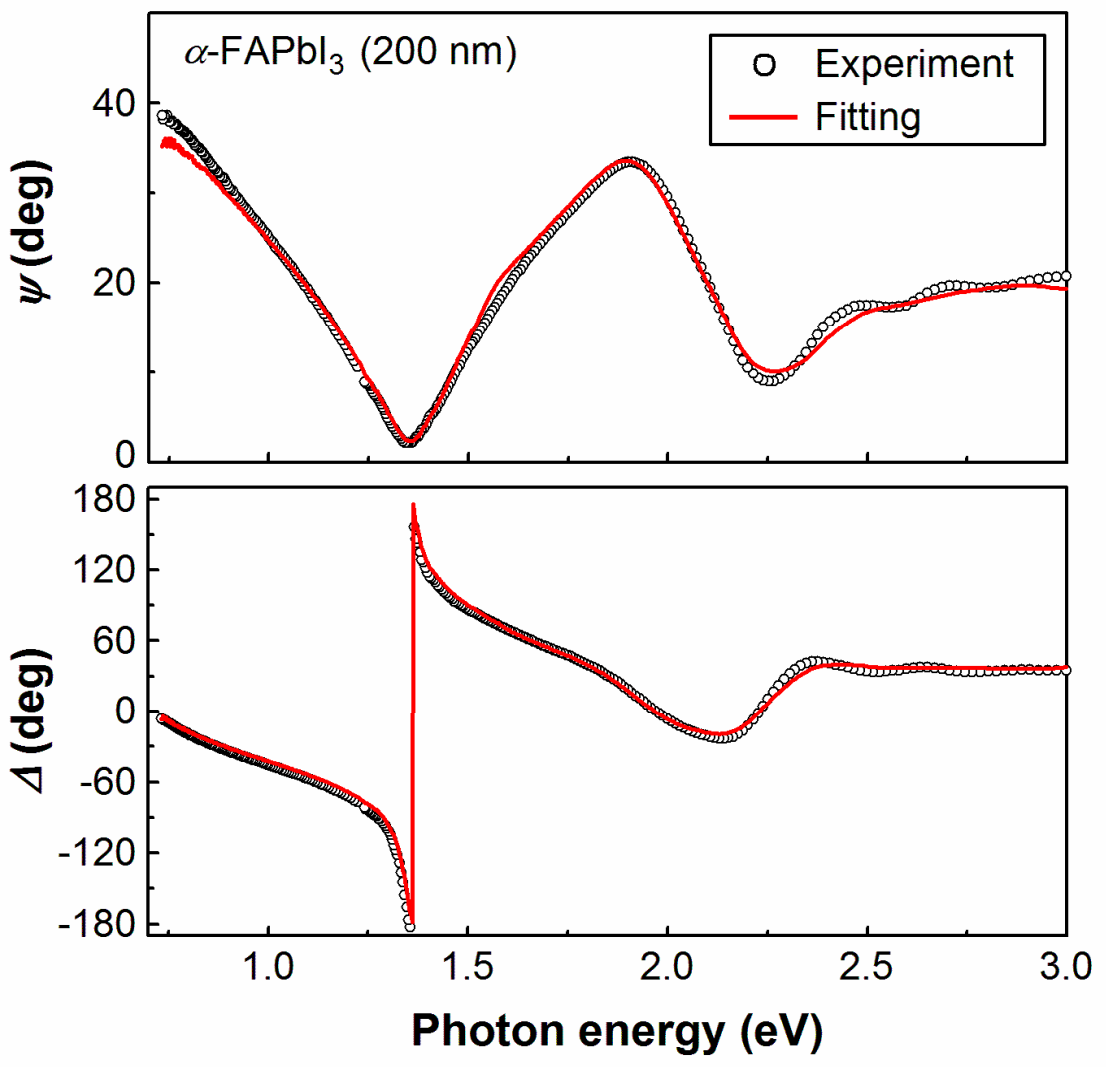

FIG. 2. ($\psi$, $\Delta$) ellipsometry spectra obtained from the $\alpha$-$FAPbI_3$ (200 nm)/$SiO_2$ (2 nm)/$c$-Si structure (open circles) and the fitting result calculated using the $\alpha$-$FAPbI_3$ dielectric function extracted from the thinner layer of 100 nm (solid lines).

## III. SE ANALYSIS

The dielectric functions of the $\alpha$-$FAPbI_3$ and $\delta$-$FAPbI_3$ samples are determined by SE. The SE measurements of the $FAPbI_3$ layers are performed in a $N_2$ ambient without exposing the samples to air by a procedure established earlier,[5] as $\alpha$-$FAPbI_3$ exhibits significant degradation in humid air.[3,46,53,55] The SE measurements are implemented at an angle of incidence of 75° using a rotating-compensator ellipsometer[71] (J.A. Woollam, M–2000XI).

The SE analyses of the $\alpha$-$FAPbI_3$ layers are performed using a global error minimization scheme,[72] in which the dielectric function is determined self-consistently using more than two samples with different layer thicknesses. For the $\alpha$-$FAPbI_3$ analysis, two samples with thicknesses of 100 and 200 nm in Fig. 1(a) are used, and two sets of ellipsometry spectra are obtained from these samples [i.e., $(\psi, \Delta)_{100nm}$ and $(\psi, \Delta)_{200nm}$]. In our analysis, the dielectric function of the bulk layer $\varepsilon_{bulk}(E)$ is extracted first from $(\psi, \Delta)_{100nm}$ using a mathematical inversion.[71] For this analysis, an optical model consisting of a surface roughness layer/$\alpha$-$FAPbI_3$ bulk layer/$SiO_2$ (2 nm)/$c$-Si substrate is assumed. For the calculation of the surface-roughness optical properties, the Bruggeman effective-medium approximation (EMA)[71] is employed. In our self-consistent SE analysis, $\varepsilon_{\text{bulk}}(E)$ obtained from $(\psi, \Delta)_{100\text{nm}}$ is applied further for the fitting analysis of $(\psi, \Delta)_{200\text{nm}}$. For this analysis, we employ an optical model with a void-rich interface layer on the substrate [i.e., surface roughness layer/$\alpha$-$FAPbI_3$ bulk layer/interface layer/$SiO_2$ (2 nm)/$c$-Si substrate], and the optical properties of the interface layer are also deduced from EMA assuming a mixture of $\alpha$-$FAPbI_3$ and voids.

Figure 2 shows the SE fitting analysis of $(\psi, \Delta)_{200\text{nm}}$. In this figure, the open circles show experimental spectra of $(\psi, \Delta)_{200\text{nm}}$ and the solid lines indicate the fitting result calculated using $\varepsilon_{\text{bulk}}(E)$ extracted from $(\psi, \Delta)_{100\text{nm}}$. As confirmed from Fig. 2, the calculated spectra show excellent fitting to the experimental spectra. In this analysis, however, the fitting quality degrades at $E \geq 3.0$ eV, most likely due to the effect of the rough surface. Thus, in the global-error-minimization analysis, the fitting is carried out in the region of $E \leq 3.0$ eV. In the subsequent analysis, we extract the final $\alpha$-$FAPbI_3$ dielectric function from $(\psi, \Delta)_{100\text{nm}}$ by adjusting the bulk layer thickness slightly (within 2% of the total thickness) so that the $\varepsilon_2$ values become zero at $E < E_g$. However, the surface roughness thickness ($d_s = 1.7 \pm 0.1$ nm) obtained from the above analysis is smaller than the value of $d_{\text{rms}} = 10.9$ nm determined from AFM [Fig. 1(b)]. Thus, the amplitude of the $\alpha$-$FAPbI_3$ dielectric function in the high energy region ($E > 3$ eV), which is excluded in the self-consistent SE analysis, could be underestimated by the effect of rough surface.[5] A similar ellipsometry analysis is performed for $\delta$-$FAPbI_3$ using the two samples in Fig. 1(a).

## IV. DFT CALCULATION

The DFT calculations of various $APbX_3$ perovskite materials are implemented using a plane-wave ultrasoft pseudopotential method (Advance/PHASE software). For the calculations, the generalized gradient approximation within the Perdew-Burke-Ernzerhof scheme (PBE)[73] has been applied without considering spin-orbit coupling (SOC). Although the SOC effect is rather significant in hybrid perovskites,[14-16,20,27,30,32] $E_g$ of hybrid perovskites such as $MAPbI_3$ and $FAPbI_3$ can still be reproduced by the simple PBE calculation without incorporating SOC. This has been interpreted by the cancellation of errors induced by (i) the underestimation of $E_g$ caused by the PBE functional and (ii) the overestimation of $E_g$ generated without considering SOC.[27,29,30] In more sophisticated calculations using the *GW* approximation[14,15,20] and a hybrid functional,[16] the experimental $E_g$ values of $MAPbI_3$ and $\alpha$-$FAPbI_3$ can be obtained by incorporating the SOC interaction. However, these high-level DFT calculations lead to the band structures, which are essentially similar to that deduced from the PBE calculation.[13,16,18] Thus, in this study, the simple PBE calculation is implemented without incorporating the SOC effect.

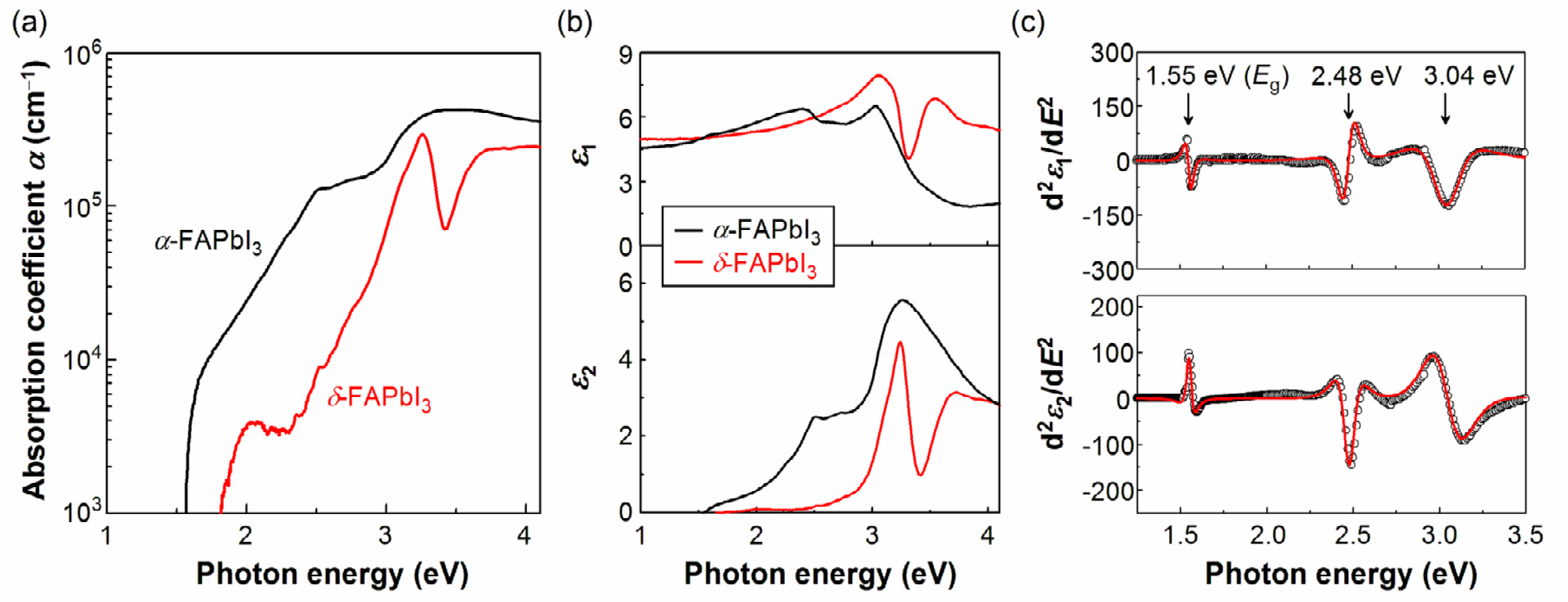

FIG. 3. (a) $\alpha$ spectra and (b) dielectric functions extracted from the $\alpha$-$FAPbI_3$ layer (100 nm) and $\delta$-$FAPbI_3$ layer (30 nm) in Fig. 1, together with (c) CP analysis of the $\alpha$-$FAPbI_3$ dielectric function. In (c), the open circles denote the experimental data and the solid lines represent the theoretical fitting performed assuming three transitions in an energy region of $E \leq 3.5$ eV. The CP energies determined from the analysis are indicated by the arrows.

For $\alpha$-$FAPbI_3$ and $MAPbBr_3$ crystals, we perform structural optimization within the cubic basis using a $6 \times 6 \times 6$ k mesh and a plane-wave cutoff energy of 500 eV until the atomic configuration converged to within 10 meV/Å. The dielectric functions are calculated based on a method developed by Kageshima et al.[74] For this calculation, we use a more dense $10 \times 10 \times 10$ k mesh to suppress distortion of calculated spectra.

## V. RESULTS

### A. Optical constants of hybrid perovskite materials

Figure 3 shows (a) the $\alpha$ spectra and (b) dielectric functions of the $\alpha$-$FAPbI_3$ and $\delta$-$FAPbI_3$ layers, together with (c) the critical point (CP) analysis of the $\alpha$-$FAPbI_3$ dielectric function. These optical data are obtained from the self-consistent SE analyses described in Sec. III. The $\alpha$ spectrum of $\alpha$-$FAPbI_3$ shows the sharp absorption onset near $E_g$ and the Urbach energy ($E_U$) determined assuming $\alpha \propto \exp(E/E_U)$ is 16 meV, supporting the suppressed tail state formation in $\alpha$-$FAPbI_3$, as observed in $MAPbI_3$ ($E_U$ = 14–15 meV)[5,28] and (FA, Cs)Pb(I, $Br)_3$ ($E_U$ = 16 meV).[56] It can be seen that the $\alpha$ value is significantly smaller in the $\delta$ phase, compared with the $\alpha$ phase, and the corresponding dielectric functions of these layers are quite different.

From the CP analysis shown in Fig. 3(c), we have further determined $E_g$ of $\alpha$-$FAPbI_3$. This analysis is carried out assuming three transitions in an energy region of $E \leq 3.5$ eV by using a procedure established previously.[5] The open circles denote the experimental data and the solid lines represent the theoretical fitting. As a result, the CP energies of $\alpha$-$FAPbI_3$ are determined to be 1.55 ± 0.01 eV, 2.48 ± 0.01 eV, and 3.04 ± 0.01 eV. The CP energy of 1.55 ± 0.01 eV corresponds to $E_g$ ($E_0$ transition) and, in earlier studies, slightly smaller values in a range of 1.43-1.53 eV have been reported.[3,45-48,53,59] The slight disagreement observed for $E_g$ could be attributed to (i) the uncertainty in the $E_g$ analysis and (ii) the underestimation of $E_g$ due to extensive roughness of samples.[5] The $E_g$ value of $MAPbI_3$ determined from a similar CP analysis is 1.61 ± 0.01 eV.[5] Thus, $E_g$ changes only slightly in $\alpha$-$FAPbI_3$, if compared with $MAPbI_3$.

Figure 4 compares (a) the $\alpha$ spectrum and (b) the dielectric function of $\alpha$-$FAPbI_3$ obtained in this study with those of $MAPbI_3$ (Ref. 5), $MAPbBr_3$ (Ref. 27) and $MAPbCl_3$ (Ref. 27) reported earlier. The optical spectra of $MAPbI_3$ in Fig. 4 correspond to those obtained in our earlier study[5] and have been extracted from a laser-evaporated $MAPbI_3$ layer using the self-consistent SE analysis, whereas the optical data of $MAPbBr_3$ and $MAPbCl_3$ have been obtained from the SE analyses of the single crystals.[27] We find that the $\alpha$ values of $\alpha$-$FAPbI_3$ are notably smaller than those of $MAPbI_3$ and, at 1.7 eV, $\alpha$ of $\alpha$-$FAPbI_3$ is half of that of $MAPbI_3$. Furthermore, although the overall $\varepsilon_2$-spectral shapes of $MAPbI_3$ and $\alpha$-$FAPbI_3$ are similar, the $\varepsilon_2$ amplitude of $\alpha$-$FAPbI_3$ is roughly half of that of $MAPbI_3$ at $E < 3$ eV. Accordingly, the replacement of the center cation has a large impact on the absorption strength. The lower $\alpha$ observed in $\alpha$-$FAPbI_3$ is consistent with the experimental fact that thicker absorber layers (~500 nm) are generally necessary for $FAPbI_3$-based perovskite solar cells,[52,57-59] compared with $MAPbI_3$ solar cells (typically ~300 nm).[1] The $\varepsilon_1$ value at a low energy ($E$ = 1 eV) also reduces from 5.26 ($MAPbI_3$) to 4.57 ($\alpha$-$FAPbI_3$) and, therefore, the refractive index also varies by the effect of the center cation.

For the change of X in $MAPbX_3$, on the other hand, the whole dielectric function shifts systematically toward higher energy and the amplitude of the $\varepsilon_2$ spectrum gradually reduces in the lighter halogen atom [Fig. 4(b)]. In addition, a high-energy transition peak observed at 3.24 eV in $MAPbI_3$ splits into two peaks in $MAPbBr_3$ and $MAPbCl_3$. The sharp absorption peaks observed near the $E_g$ regions of $MAPbBr_3$ and $MAPbCl_3$ are caused by strong excitonic transitions[75] and the excitonic feature becomes stronger in the lighter halogen atom. The results of Fig. 4 show clearly that the A-site cation varies the oscillator strength of the optical transition, while the X-site halogen atom determines the transition energy.

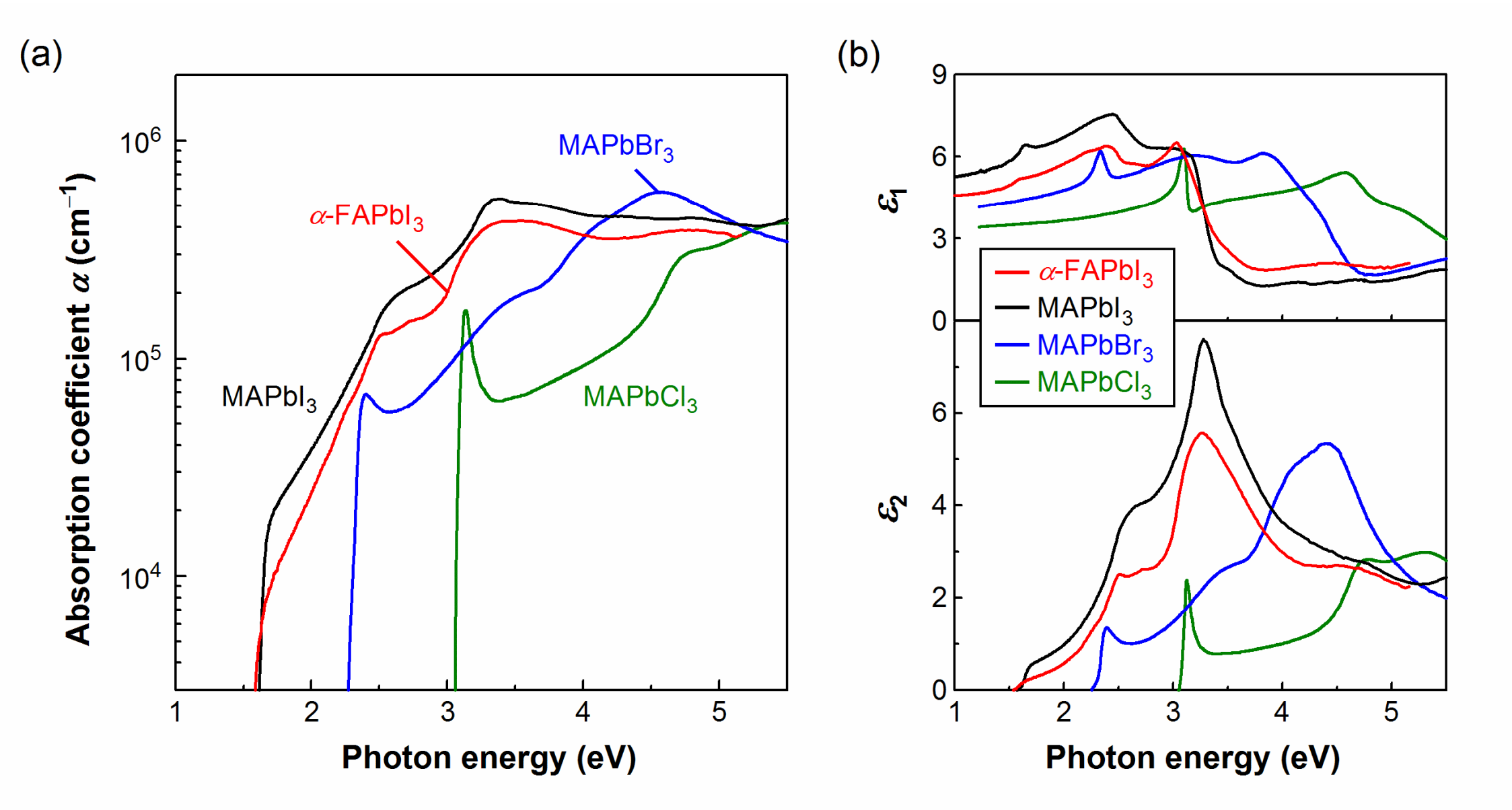

FIG. 4. (a) $\alpha$ spectra and (b) dielectric functions of $\alpha$-$FAPbI_3$, $MAPbI_3$, $MAPbBr_3$ and $MAPbCl_3$. The optical data of $\alpha$-$FAPbI_3$ in Fig. 3 and reported optical data of $MAPbI_3$ (Ref. 5), $MAPbBr_3$ (Ref. 27) and $MAPbCl_3$ (Ref. 27) are shown. The results of $\alpha$-$FAPbI_3$ and $MAPbI_3$ are obtained from the polycrystalline materials, whereas those of $MAPbBr_3$ and $MAPbCl_3$ are extracted from the single crystals.

The low $\alpha$ values observed in $\alpha$-$FAPbI_3$ are justified further from external quantum efficiency (EQE) analysis. In particular, we analyze the EQE spectrum of a reported high-efficiency $FAPbI_3$ solar cell fabricated by a solution-based process[52] using the $\alpha$-$FAPbI_3$ dielectric function shown in Fig. 4(b). This $FAPbI_3$ solar cell has a structure of glass/$SnO_2$:F/compact $TiO_2$/mesoporous $TiO_2$ – $\alpha$-$FAPbI_3$/$\alpha$-$FAPbI_3$/polytriarylamine (PTAA)/Au.[52] For the optical constants of the component layers, we employ reported data,[5,76] although the optical constants of spiro-OMeTAD [2,2',7,7'-tetrakis-($N$,$N$-di-$p$-methoxyphenylamine) 9,9'-spirobifluorene][5] are used for the PTAA hole transport layer (HTL), as the optical constants of PTAA have not been reported and the absorption properties of PTAA[77] are similar to those of spiro-OMeTAD. The optical response within the mesoporous mixed-phase layer is expressed by two separate flat layers of $TiO_2$ and $\alpha$-$FAPbI_3$ assuming a porosity of 60% for the $TiO_2$ layer.[5] From these assumptions, we construct an optical model of glass/$SnO_2$:F (600 nm)/$TiO_2$ (120 nm)/$\alpha$-$FAPbI_3$ (590 nm)/spiro-OMeTAD (50 nm)/Au for the solar cell. The actual EQE calculation is implemented based on a reported procedure in which the absorptance of each component layer in solar cells is deduced from the optical admittance method using an experimental reflectance ($R$) spectrum.[78] However, since the $R$ spectrum of the perovskite solar cell has not been reported, we perform the EQE calculation assuming a fixed $R$ value of 10%, as observed in $MAPbI_3$ solar cells having similar structures ($R$ = 2–10 %).[2] The EQE spectrum of the solar cell is then determined assuming 100% collection of photocarriers generated within the $\alpha$-$FAPbI_3$ absorber layer.

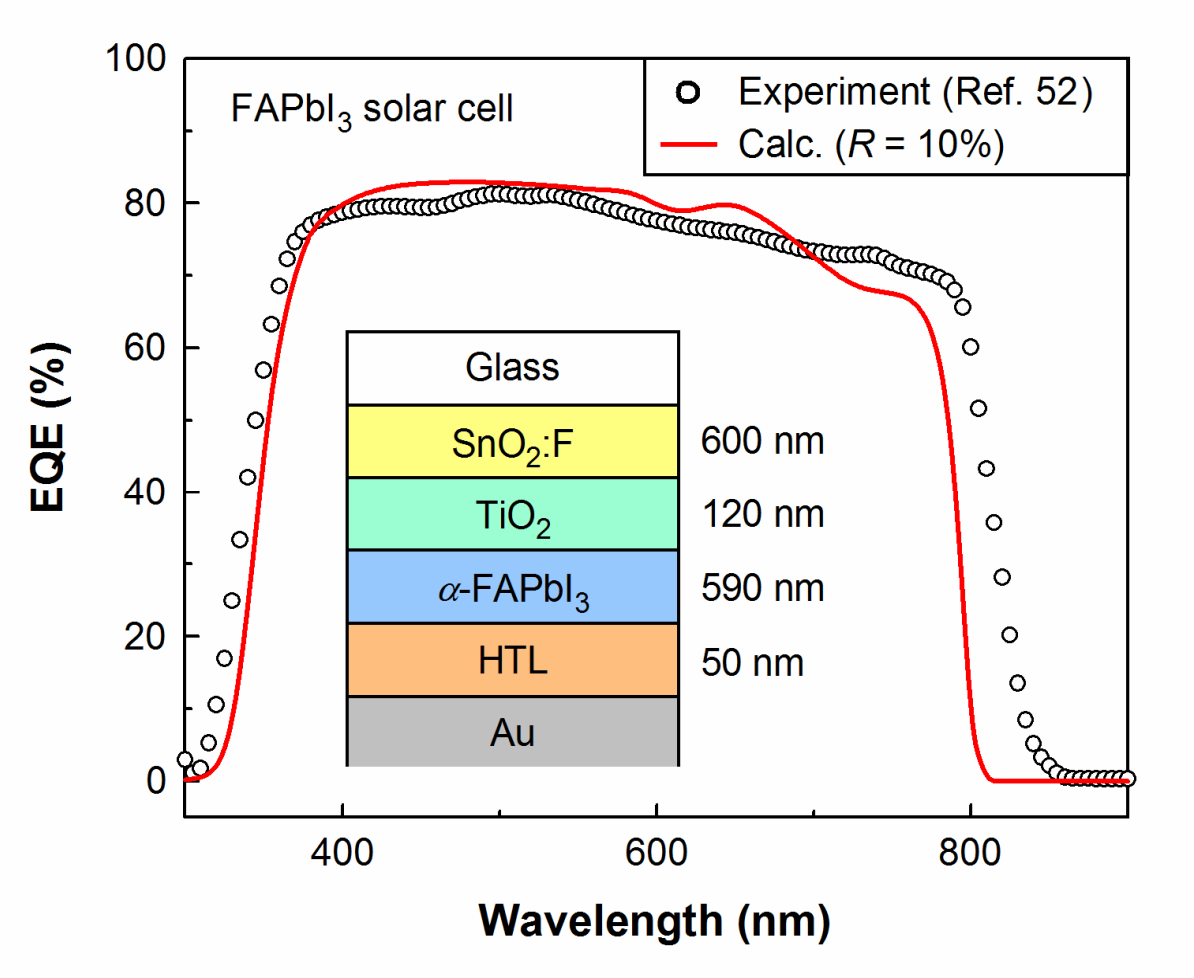

FIG. 5. EQE analysis of a $FAPbI_3$ hybrid perovskite solar cell using the $\alpha$-$FAPbI_3$ optical constants shown in Fig. 4. The $FAPbI_3$ hybrid-perovskite solar cell fabricated by a solution process[52] has a structure of glass/$SnO_2$:F/compact $TiO_2$/mesoporous $TiO_2$ – $\alpha$-$FAPbI_3$/uniform $\alpha$-$FAPbI_3$/polytriarylamine (PTAA)/Au. The open circles show the experimental data of Ref. 52, whereas the solid line represents the EQE spectrum calculated using an optical model consisting of glass/$SnO_2$:F (600 nm)/$TiO_2$ (120 nm)/$\alpha$-$FAPbI_3$ (590 nm)/HTL (50 nm)/Au. In the EQE calculation, $R$ of the solar cell is assumed to be 10% in the whole wavelength region and the optical constants of spiro-OMeTAD are used for the HTL.

Figure 5 shows the experimental EQE spectrum of the $FAPbI_3$ hybrid perovskite solar cell[52] (open circles) and the simulated EQE spectrum obtained from the above calculation procedure assuming $R$ = 10% (solid line). As confirmed from Fig. 5, the overall EQE response in the solar cell is reproduced quite well in our simulation. However, the absorption onset of α-$FAPbI_3$ observed in the experimental EQE spectrum is ~840 nm and shows a slight red shift, compared with the EQE calculation result (~800 nm). This could be attributed to (i) the light scattering in the solar cell, which is not modeled in our EQE analysis or (ii) the tail-state absorption in the solar cell.

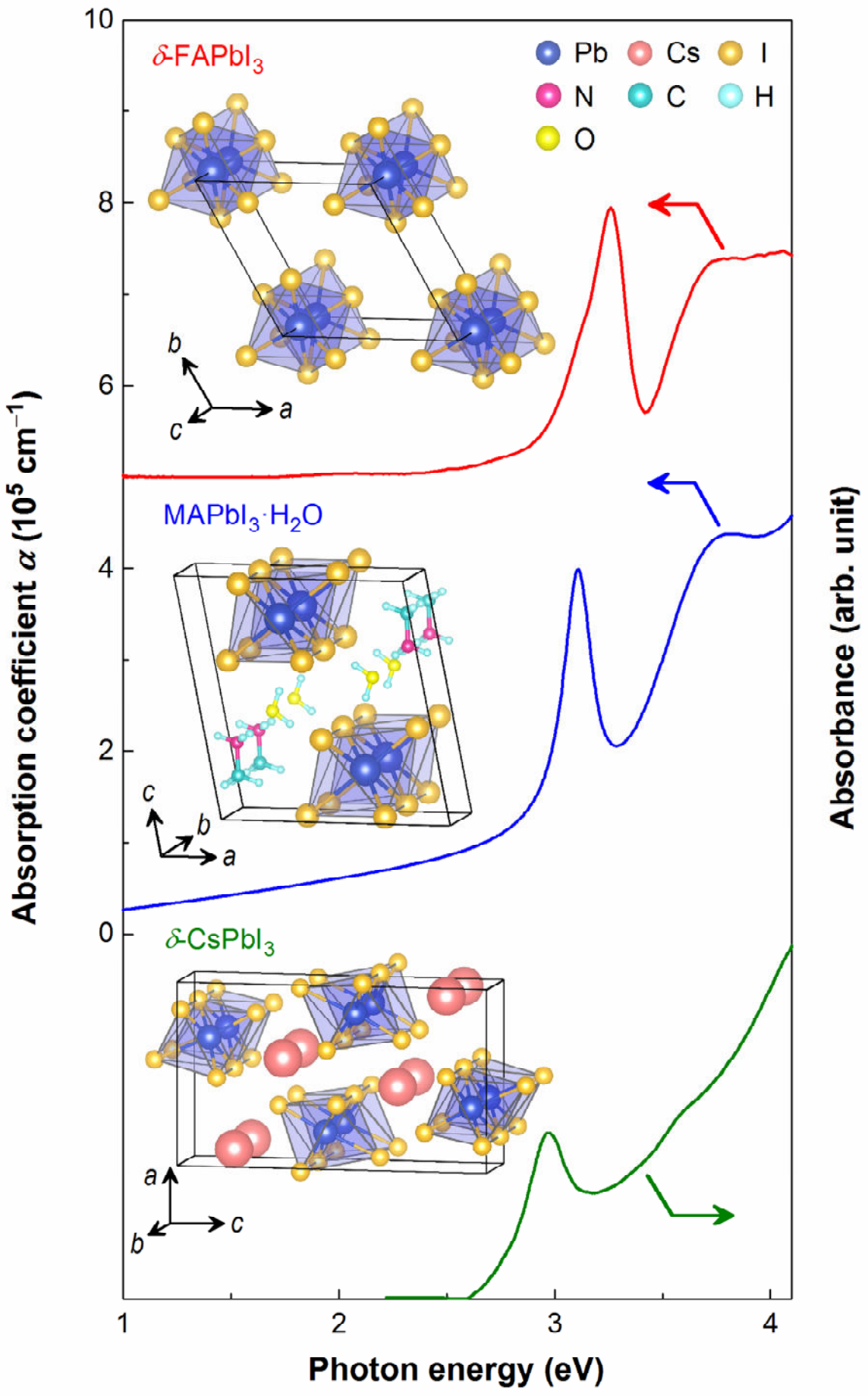

FIG. 6. Absorption spectra of δ-$FAPbI_3$, $MAPbI_3 \cdot H_2O$ and δ-$CsPbI_3$. The α spectrum of δ-$FAPbI_3$ corresponds to that shown in Fig. 3(a), whereas the absorption spectra of $MAPbI_3 \cdot H_2O$ and δ-$CsPbI_3$ are taken from Ref. 43 and Ref. 79, respectively. The α spectrum of δ-$FAPbI_3$ is shifted by $5 \times 10^5$ $cm^{-1}$ for clarity. The insets show the crystal structures reported for δ-$FAPbI_3$ (Ref. 3), $MAPbI_3 \cdot H_2O$ (Ref. 80) and δ-$CsPbI_3$ (Ref. 3). In the structure of δ-$FAPbI_3$, the FA cations are not shown due to the uncertainty of the exact configuration.

In Fig. 6, on the other hand, the α spectrum of δ-$FAPbI_3$ is compared with the absorption spectra of $MAPbI_3{\cdot}H_2O$ (Ref. 43) and orthorhombic $CsPbI_3$ ($\delta$-$CsPbI_3$),[79] which also have one-dimensional crystal structures. It can be seen that all the $PbI_6$-based compounds in Fig. 6 exhibit similar absorption peaks at $E$ = 3.0-3.3 eV with weak shoulder peaks at 3.6-3.8 eV. The sharp peak at 3.1 eV in $MAPbI_3{\cdot}H_2O$ cannot be reproduced in the DFT calculation within the PBE[11] and has been attributed to the excitonic transition.[43] As confirmed from the insets of Fig. 6, the crystal structure of $\delta$-$FAPbI_3$ (Ref. 3) is slightly different from those of $MAPbI_3{\cdot}H_2O$ (Ref. 80) and $\delta$-$CsPbI_3$ (Ref. 3), although the absorption feature of $\delta$-$FAPbI_3$ is quite similar to those of $MAPbI_3{\cdot}H_2O$ and $\delta$-$CsPbI_3$.

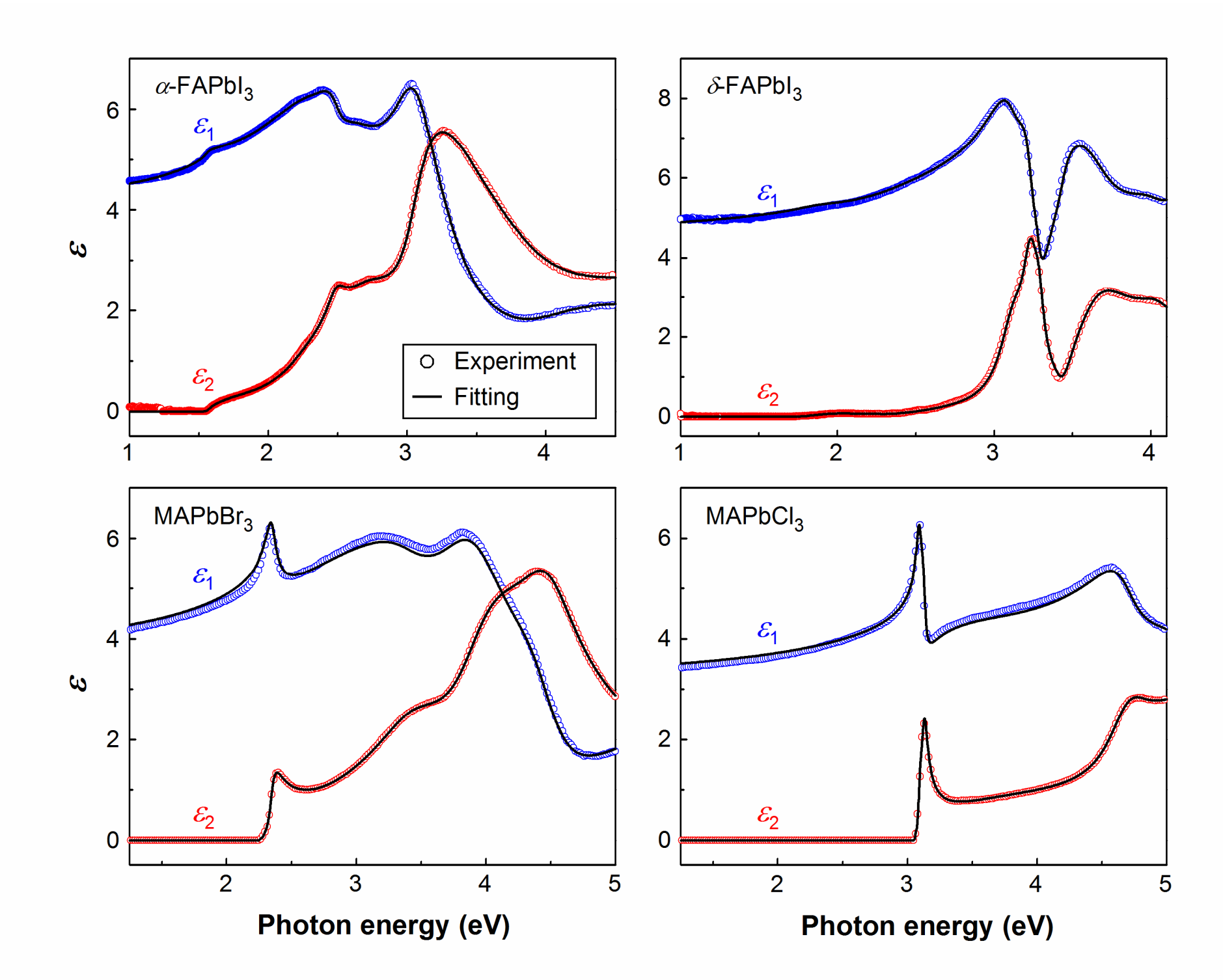

FIG. 7. Parameterization of the dielectric functions of $\alpha$-$FAPbI_3$, $\delta$-$FAPbI_3$, $MAPbBr_3$ and $MAPbCl_3$ using Tauc-Lorentz transition peaks. The open circles show the experimental dielectric functions of $\alpha$-$FAPbI_3$ and $\delta$-$FAPbI_3$ shown in Fig. 3(b) and $MAPbBr_3$ (Ref. 27) and $MAPbCl_3$ (Ref. 27) shown in Fig. 4(b), while the solid lines indicate the fitting results. The model parameters are summarized in Table I.

The dielectric functions of $\alpha$-$FAPbI_3$, $MAPbBr_3$ and $MAPbCl_3$ in Fig. 4(b) can be parameterized by assuming several Tauc-Lorentz (TL) transitions.[81] In the TL model, a dielectric function peak is modeled by a total of five parameters: the amplitude parameter ($A$), broadening parameters ($C$), peak transition energy ($E_p$), Tauc optical gap ($E_T$) and the $\varepsilon_1$ contribution at high energy [$\varepsilon_1(\infty)$]. Figure 7 summarizes the results of dielectric function modeling performed for $FAPbI_3$, $MAPbBr_3$ and $MAPbCl_3$ using the TL peaks. In this figure, the open circles denote the optical data shown in Fig. 4, whereas the solid lines indicate the calculation results. The TL parameters for $FAPbI_3$, $MAPbBr_3$ and $MAPbCl_3$ are summarized in Table 1. From these model parameters, the $\alpha$ spectra of these materials, shown in Fig. 4(a), can be expressed. Similar parameterization has already been performed for $MAPbI_3$ in our earlier study.[5]

**TABLE I.** Tauc-Lorentz parameters extracted from the dielectric function modeling shown in Fig. 7.

| Material | Peak | $E_p$ (eV) | $A$ (eV) | $C$ (eV) | $E_T$ (eV) | $\varepsilon_1(\infty)$ |
|---|---|---|---|---|---|---|
| $\alpha$-$FAPbI_3$ | 1 | 1.557 | 10.018 | 0.099 | 1.543 | 1.462 |
| | 2 | 1.566 | 8.849 | 0.498 | 1.521 | 0 |
| | 3 | 2.344 | 9.041 | 0.794 | 1.618 | 0 |
| | 4 | 2.486 | 1.028 | 0.171 | 1.740 | 0 |
| | 5 | 2.712 | 20.435 | 0.755 | 2.093 | 0 |
| | 6 | 3.083 | 66.516 | 0.417 | 2.726 | 0 |
| | 7 | 3.190 | 98.353 | 0.657 | 2.918 | 0 |
| | 8 | 3.897 | 131.316 | 3.018 | 3.882 | 0 |
| $\delta$-$FAPbI_3$ | 1 | 1.893 | 1.573 | 0.404 | 1.667 | 2.455 |
| | 2 | 2.291 | 15.657 | 1.898 | 2.288 | 0 |
| | 3 | 3.127 | 2.182 | 0.232 | 1.690 | 0 |
| | 4 | 3.232 | 1.563 | 0.118 | 1.803 | 0 |
| | 5 | 3.285 | 0.411 | 0.078 | 1.695 | 0 |
| | 6 | 3.439 | 367.887 | 0.233 | 3.411 | 0 |
| | 7 | 3.611 | 72.513 | 0.366 | 3.398 | 0 |
| | 8 | 4.021 | 3.777 | 0.291 | 3.305 | 0 |
| | 9 | 4.638 | 776.547 | 0.202 | 4.388 | 0 |
| $MAPbBr_3$ | 1 | 2.258 | 14.194 | 0.022 | 2.245 | 0.011 |
| | 2 | 2.298 | 50.951 | 0.110 | 2.257 | 0 |
| | 3 | 2.343 | 140.251 | 0.090 | 2.305 | 0 |
| | 4 | 2.425 | 48.060 | 2.641 | 2.382 | 0 |
| | 5 | 3.416 | 14.262 | 1.002 | 2.412 | 0 |
| | 6 | 3.949 | 78.865 | 0.663 | 3.489 | 0 |
| | 7 | 4.466 | 6.204 | 0.620 | 2.428 | 0 |
| | 8 | 8.469 | 41.650 | 0.070 | 2.905 | 0 |

**TABLE I.** (*Continued.*)

| Material | Peak | $E_p$ (eV) | $A$ (eV) | $C$ (eV) | $E_g$ (eV) | $\varepsilon_1(\infty)$ |
|---|---|---|---|---|---|---|
| $MAPbCl_3$ | 1 | 3.065 | 24.330 | 0.012 | 3.045 | 0.076 |
| | 2 | 3.096 | 177.556 | 0.046 | 3.043 | 0 |
| | 3 | 3.125 | 219.273 | 0.052 | 3.071 | 0 |
| | 4 | 3.199 | 34.078 | 1.516 | 3.099 | 0 |
| | 5 | 4.695 | 10.013 | 0.397 | 3.733 | 0 |
| | 6 | 5.279 | 17.475 | 1.421 | 3.051 | 0 |
| | 7 | 9.718 | 209.001 | 0.053 | 6.544 | 0 |

### B. Optical transitions in $\alpha$-$FAPbI_3$ and $MAPbBr_3$

To understand the optical transitions in $\alpha$-$FAPbI_3$ and $MAPbBr_3$ crystals, the DFT calculations are implemented assuming pseudocubic structures. Figure 8(a) shows the $\alpha$-$FAPbI_3$ and $MAPbBr_3$ crystal structures determined from the DFT. In Fig. 8(a), $a$, $b$, and $c$ indicate the axes of the unit cell. From the structural optimization, we obtain lattice parameters of $a$ = 6.416 Å, $b$ = 6.236 Å, and $c$ = 6.353 Å with $\alpha = \beta = \gamma = 90.00°$ ($\alpha$-$FAPbI_3$) and $a$ = 5.934 Å, $b$ = 5.919 Å, and $c$ = 5.936 Å with $\alpha = \gamma = 90.00°$ and $\beta = 89.94°$ ($MAPbBr_3$). In the DFT result for $\alpha$-$FAPbI_3$, the lattice parameter of $a$ is expanded slightly, compared with $b$ and $c$, due to the steric effect of $FA^+$, while the lattice parameters are almost the same in $MAPbBr_3$. The average lattice parameters obtained from the calculations (6.335 Å in $\alpha$-$FAPbI_3$ and 5.930 Å in $MAPbBr_3$) show excellent agreement with the experimental results reported for $\alpha$-$FAPbI_3$ (6.362 Å in Ref. 50) and $MAPbBr_3$ (5.94 Å in Ref. 10). In $\alpha$-$FAPbI_3$, the introduction of the large center cation leads to the formation of collinear I-Pb-I bonds, which are more closer to 180°, if compared with $MAPbI_3$.[5]

In many earlier studies, $\alpha$-$FAPbI_3$ is reported to have a trigonal symmetry.[3,45,47-49] Nevertheless, a recent neutron-powder diffraction study[50] confirms that $\alpha$-$FAPbI_3$ has a simpler cubic symmetry with the N-C-N plane parallel to the *a*-*b* plane (or the line connecting two N atoms of $FA^+$ is parallel to the *a* axis). This experimental result is consistent with our DFT result shown in Fig. 8(a). The configuration of $MA^+$ in $MAPbBr_3$ is essentially similar to that in $MAPbI_3$ (Ref. 5) and the C–N bond of $MA^+$ is almost parallel to the *a* axis.

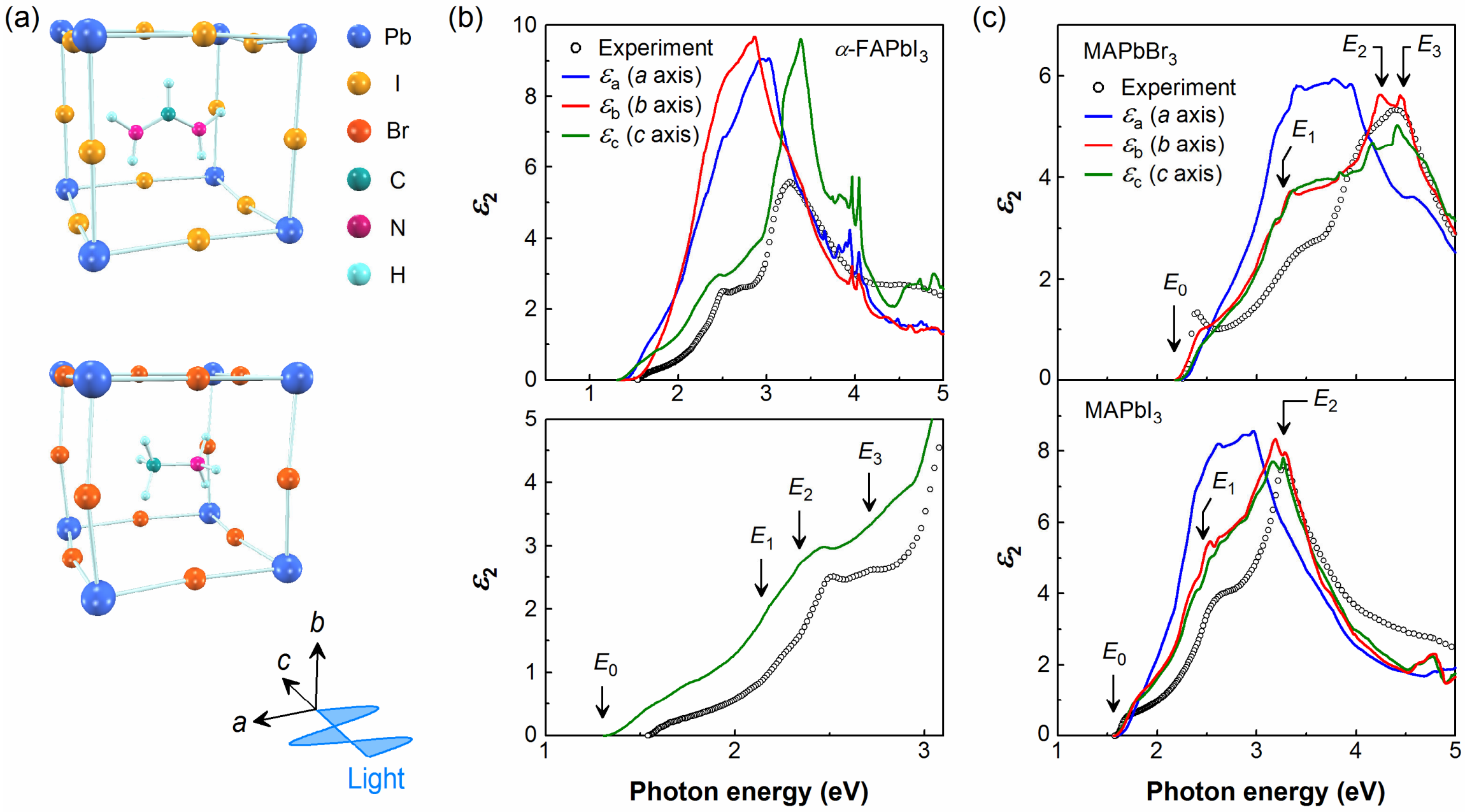

FIG. 8. (a) Pseudocubic crystal structures of $\alpha$-$FAPbI_3$ and $MAPbBr_3$ obtained from the DFT calculations, and $\varepsilon_2$ spectra of (b) $\alpha$-$FAPbI_3$ and (c) $MAPbBr_3$ and $MAPbI_3$, obtained from the experiments (open circles) and DFT calculations (solid lines). In (b) and (c), the experimental results correspond to those shown in Fig. 4(b) ($MAPbBr_3$: Ref. 27, $MAPbI_3$: Ref. 5), and $\varepsilon_a$, $\varepsilon_b$ and $\varepsilon_c$ represent the $\varepsilon_2$ spectra calculated assuming the light polarization along the *a*, *b*, and *c* axes shown in (a), respectively. The arrows indicate the transition energies determined by the DFT analyses. In (b), the $\varepsilon_2$ spectra in two different energy regions are shown for clarity. In (c), the DFT calculation result for $MAPbBr_3$ has been shifted toward higher energy by 0.22 eV to obtain better matching with the experimental spectrum, whereas the results of $MAPbI_3$ are adopted from our earlier study[5].

In Fig. 8(b), the experimental $\varepsilon_2$ spectrum of $\alpha$-$FAPbI_3$ is compared with the $\varepsilon_2$ spectra that are calculated for the different polarization states with directions parallel to the $a$, $b$, and $c$ axes in Fig. 8(a). The DFT calculation reveals the highly anisotropic optical behavior of $\alpha$-$FAPbI_3$ and the $\varepsilon_2$ spectra for the $a$ and $b$ axes ($\varepsilon_a$ and $\varepsilon_b$) exhibit quite different shapes from that for the $c$ axis ($\varepsilon_c$). In Fig. 8(b), the $\varepsilon_2$ spectra in two different energy regions are shown for $\varepsilon_c$. Quite remarkably, $\varepsilon_c$ reproduces all the fine absorption features observed experimentally at $E < 3$ eV and shows the excellent overall agreement with the experimental result. At higher energies ($E > 3$ eV), however, the amplitude of the experimental $\varepsilon_2$ spectrum is much lower than that of the calculation ($\varepsilon_c$). The disagreement in the high energy region may originate from SE analysis errors induced by the rough surface structures of the $\alpha$-$FAPbI_3$ samples, as mentioned in Sec. III. In Fig. 8(b), the arrows represent the optical transitions in the Brillouin zone, determined from the polarization-dependent DFT analysis described below.

In Fig. 8(c), the $\varepsilon_2$ spectra of $MAPbBr_3$ calculated from the structure of Fig. 8(a) are compared with the experimental $\varepsilon_2$ spectrum of Ref. 27. The experimental and DFT data of $MAPbI_3$ in Fig. 8(c) are taken from our previous study,[5] in which the DFT calculation was performed in a similar way. For $MAPbBr_3$, all the calculated spectra have been shifted toward higher energy by 0.22 eV to match the DFT result with the experimental result. As confirmed from Fig. 8(c), $MAPbBr_3$ and $MAPbI_3$ show the strong anisotropic behavior, and the experimental $\varepsilon_2$ spectra of $MAPbBr_3$ and $MAPbI_3$ are reproduced surprisingly well in the calculated $\varepsilon_2$ spectra of $\varepsilon_b$ and $\varepsilon_c$. The $\varepsilon_a$ spectrum of $MAPbBr_3$ is similar to that reported in a previous DFT study.[63] The arrows for $MAPbBr_3$ in Fig. 8(c) show the result of our optical transition analysis described below. In our study, the $E_0$ ($E_g$), $E_1$ and $E_2$ transitions in $MAPbI_3$ are attributed to the non-excitonic direct transitions at the $R$, $M$ and $X$ points in the pseudocubic Brillouin zone.[5] For the assignment of the interband transitions in $MAPbI_3$, a similar result has also been reported.[27]

The DFT calculations in this study are performed without incorporating electron-hole interactions (or exciton formation). Thus, the band-edge excitonic transition observed in $MAPbBr_3$ at 2.4 eV cannot be reproduced in our DFT calculation and, consequently, the calculated $\varepsilon_2$ becomes lower than the experimental $\varepsilon_2$ near the $E_g$ region. For $\alpha$-$FAPbI_3$, the calculated $\varepsilon_2$ values are larger than the experimental $\varepsilon_2$ values, indicating the negligible contribution of excitons in the light absorption process. Non-excitonic nature of $MAPbI_3$ has already been confirmed earlier.[5,19,23]

We further calculate the band structure and density of states (DOS) of $\alpha$-$FAPbI_3$ [Fig. 9(a)] and $MAPbBr_3$ [Fig. 9(b)]. For $MAPbBr_3$, however, the energy positions of all the conduction bands are shifted upward by 0.22 eV (scissors operation).[82] The high symmetry points in the Brillouin zone defined by the reciprocal lattices ($a^* = 2\pi/a$, $b^* =$ $2\pi/b$, $c^* = 2\pi/c$) are shown in the inset of Fig. 9(b). Since the assumed pseudocubic structures are distorted, the energy positions of $M_{1-3}$ and $X_{1-3}$ in the Brillouin zone differ slightly. In Figs. 9(a) and 9(b), $V_j$ and $C_j$ denote the $j$th valence and conduction bands from the valence band maximum (VBM) and the conduction band minimum (CBM), respectively. The $V_1$ mainly consists of the I 5$p$ state in $\alpha$-$FAPbI_3$ and the Br 4$p$ state in $MAPbBr_3$, whereas $C_1$ is dominated by the Pb 6$p$ in both $\alpha$-$FAPbI_3$ and $MAPbBr_3$. It can be seen that the shapes of $C_1$ and $V_j$ ($j \leq 4$) in $\alpha$-$FAPbI_3$ and $MAPbBr_3$ are quite similar, although the energy positions of these bands are different. In the case of $\alpha$-$FAPbI_3$, there is an additional conduction band at the energy position of ~2.9 eV in Fig. 9(a). This band is due to the π-state of the $sp^2$ C atom in $FA^+$ and the DOS in this energy region increases sharply by the presence of the C 2$p$ π-state.

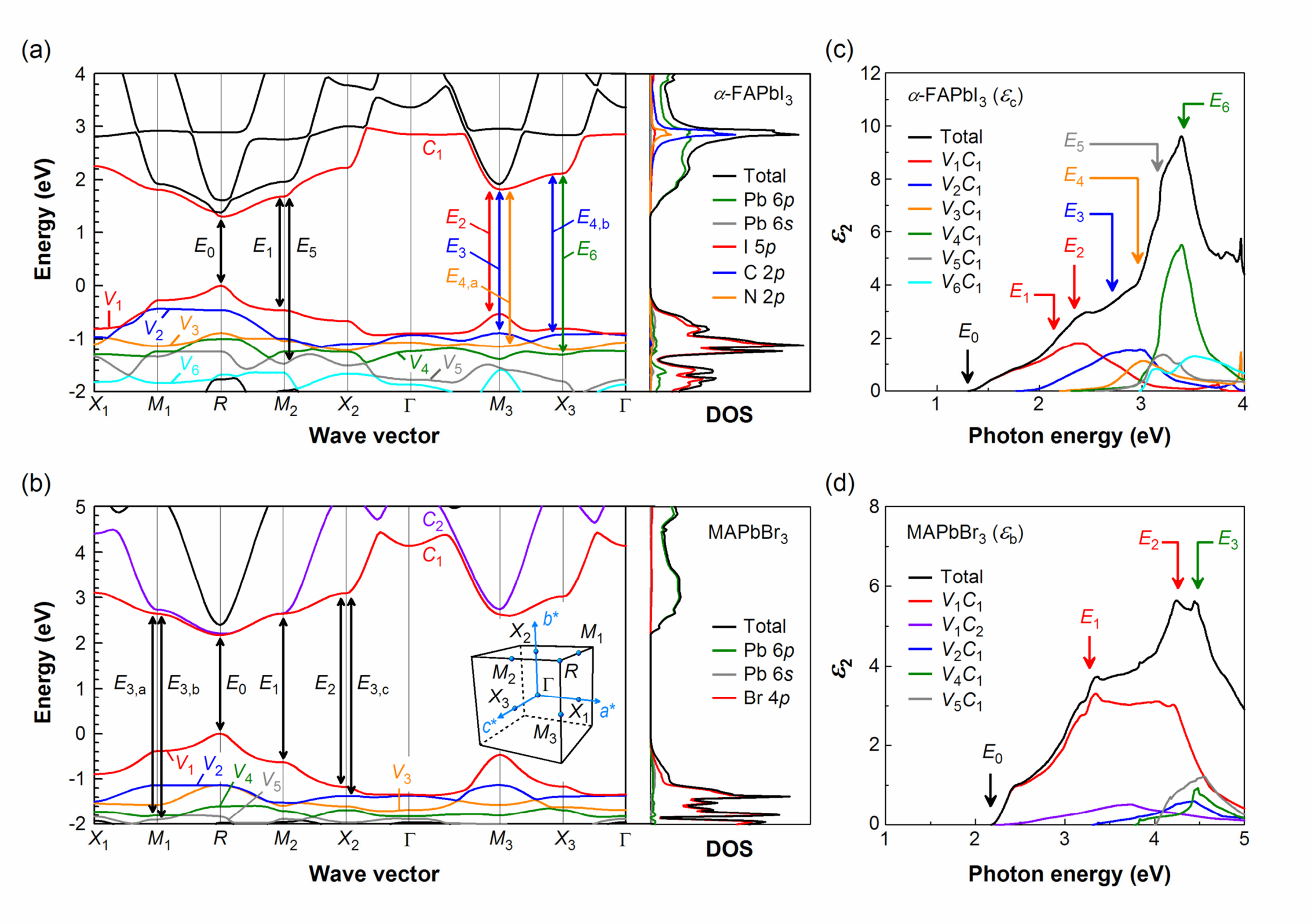

FIG. 9. Band structure and DOS of (a) $\alpha$-$FAPbI_3$ and (b) $MAPbBr_3$ pseudocubic crystals, together with the contributions of various interband transitions to (c) the $\varepsilon_c$ spectrum of $\alpha$-$FAPbI_3$ and (d) the $\varepsilon_b$ spectrum of $MAPbBr_3$. In (a) and (b), $V_j$ and $C_j$ denote the $j$th valence and conduction bands from VBM and CBM, respectively. The optical transitions determined by the polarization-dependent DFT analyses are indicated by arrows, and the partial DOS distributions are also indicated. In (a), the transition energies of $E_{4,a}$ at the $M_3$ point and $E_{4,b}$ at the $X_3$ point are almost identical. In (b), all the conduction bands have been shifted upward by 0.22 eV to improve the agreement with the experimental result. The inset shows the high symmetry points in the Brillouin zone defined by the reciprocal lattices ($a^* = 2\pi/a$, $b^* = 2\pi/b$, $c^* = 2\pi/c$) of the assumed pseudocubic structure. The transition energies of $E_{3,a}$ ($M_1$ point), $E_{3,b}$ ($M_1$ point) and $E_{3,c}$ ($X_2$ point) are almost identical. In (c) and (d), the $\varepsilon_2$ spectra denoted as "Total" correspond to $\varepsilon_c$ of $\alpha$-$FAPbI_3$ and $\varepsilon_b$ of $MAPbBr_3$ in Fig. 8, and $V_jC_k$ denotes the interband transition from the $j$th valence band to the $k$th conduction band. The transition energies ($E_{0-6}$) determined from the DFT analysis are also indicated. In (c), only the transitions with a peak amplitude of $\varepsilon_2 > 1.0$ are shown for clarity. In (d), only the transitions with a peak amplitude of $\varepsilon_2 > 0.5$ are shown for clarity.

To determine the optical transitions in $\alpha$-$FAPbI_3$ and $MAPbBr_3$, the contribution of each interband transition to the $\varepsilon_2$ spectrum is calculated for $\varepsilon_c$ of $\alpha$-$FAPbI_3$ and $\varepsilon_b$ of $MAPbBr_3$, and the results are shown in Figs. 9(c) and 9(d). In these figures, the $\varepsilon_2$ spectra denoted as "Total" correspond to those in Fig. 8, and $V_jC_k$ denotes the $\varepsilon_2$ contribution induced by the transition from the $j$th valence band to the $k$th conduction band. In Figs. 9(c) and 9(d), only the transitions with amplitudes of $\varepsilon_2 > 1.0$ ($\alpha$-$FAPbI_3$) and $\varepsilon_2 > 0.5$ ($MAPbBr_3$) are shown for clarity. Based on the band structure and the $\varepsilon_2$ contributions of $V_jC_k$, we select the optical transitions at high symmetry points that satisfy van Hove singularities[70] in $k$ space: i.e., $\nabla_{\mathbf{k}} [E_c(\mathbf{k}) - E_v(\mathbf{k})] = 0$, where $E_c(\mathbf{k})$ and $E_v(\mathbf{k})$ show the energies of the conduction and valence bands, respectively. The optical transitions determined from the DFT analyses are denoted by the arrows in Fig. 9. In Fig. 8, the transition energies of $E_{0\text{-}3}$ determined by the above procedure have been shown.

As confirmed from Figs. 9(a) and 9(b), $\alpha$-$FAPbI_3$ and $MAPbBr_3$ are direct transition semiconductors with the $E_0$ ($E_g$) transitions at the $R$ point (cubic symmetry).[27,63-65] Our analyses further show that the interband transitions in $\alpha$-$FAPbI_3$ are dominated by the transitions to $C_1$, and $C_j$ ($j \geq 2$) do not involve in the visible-light absorption. As a result, in $\alpha$-$FAPbI_3$, the interband transitions at low energies ($E < 3.2$ eV) are characterized by $V_j$ ($j \leq 5$) $\rightarrow$ $C_1$ transitions that occur predominantly at the $M$ point ($E_{1\text{-}5}$ transitions), whereas the intense $\varepsilon_2$ peak at 3.4 eV is attributed to the $V_4C_1$ transition at the $X_3$ point ($E_6$ transition). It should be noted that the $E_1$ and $E_2$ peaks of $\alpha$-$FAPbI_3$ originate from the same interband transition of $V_1C_1$ at the $M$ point. However, the energy positions of $V_1C_1$ at the $M_2$ point ($E_1$ transition) and $M_3$ point ($E_2$ transition) are different. This can be interpreted by the large difference in the lattice parameters ($a$ = 6.416 Å and $b$ = 6.236 Å) and the resulting reciprocal lattices [$a^* \rightarrow M_3$ and $b^* \rightarrow M_2$ in the Brillouin zone of Fig. 9(b)]. Accordingly, the $E_1$ and $E_2$ peaks observed in the experimental spectrum [Fig. 8(b)] support the rather significant distortion of $\alpha$-$FAPbI_3$ crystals.

In $MAPbBr_3$, on the other hand, the optical transition in the visible region occurs predominantly by the $V_1C_1$ transition [Fig. 9(d)], while the $V_1C_2$ transition shows a weak contribution. The overall optical transitions in $MAPbBr_3$ are quite similar to those of $MAPbI_3$ (Ref. 5) and the $E_1$, $E_2$ and $E_3$ peaks that appear in $\varepsilon_b$ of $MAPbBr_3$ are attributed primarily to the transitions at the $M_2$ ($V_1C_1$), $X_2$ ($V_1C_1$) and $M_1$ ($V_4C_1$ and $V_5C_1$) points, respectively. In contrast to $MAPbBr_3$, $MAPbI_3$ shows a single transition peak in the high energy region [$E_2$ transition in Fig. 8(c)], as the transition energies at the $X_2$ ($V_1C_1$) and $M_1$ ($V_4C_1$) points are almost the same in $MAPbI_3$.[5] For $MAPbBr_3$, the optical transition analysis has been made by the self-consistent *GW*[27] and the PBE,[63] but the proposed assignments of the transition peaks are slightly different from our result shown in Fig. 9(b). The DFT results in Fig. 9 indicate clearly that the interband transitions in the visible region occur predominantly from the valence state of the X atom to the Pb-derived conduction state (i.e., $V_j \rightarrow C_1$).

The $E_g$ of $\alpha$-$FAPbI_3$ obtained from the DFT calculation [1.30 eV in Fig. 9(a)] is slightly smaller than that of $MAPbI_3$ (1.56 eV).[5] It has been confirmed previously that $E_g$ of $APbI_3$ is determined primarily by the I–Pb–I bond angle of the $PbI_3^-$ cage.[17,20,44] More specifically, when the cation size is larger, the I–Pb–I bond angle becomes collinear (~180°), and resulting $E_g$ reduces slightly.[17,20,44] The decrease in $E_g$ at a larger I-Pb-I angle has been explained by the anti-bonding character of the Pb–I bond.[17,44] Consequently, $E_g$ of $APbI_3$ increases with the order of $FA^+ < MA^+ < Cs^+$.[17] As mentioned earlier, however, the experimental $E_g$ values of $\alpha$-$FAPbI_3$ (1.55 eV) and $MAPbI_3$ (1.61 eV) are rather similar.

Figure 10 shows the $\alpha$ spectra of $FAPbI_3$ calculated by the PBE without SOC (this study) and the *GW* approximation with SOC,[20] together with the experimental spectrum of Fig. 4(a). The $\alpha$ spectrum of the PBE without SOC, calculated from $\varepsilon_c$ of Fig. 8(b), shows similar trend with the experimental result. Although the anisotropic optical properties of $FAPbI_3$ have not been reported in earlier DFT studies,[20,65] the $\alpha$ spectrum derived from the simple PBE shows quite good agreement with that calculated from the *GW* method with SOC. In particular, the peak energies calculated by these DFT methods are quite consistent. However, the $\alpha$ spectra obtained from these calculations show different shapes near the $E_g$ region. As known well,[14,15,20,30] when the DFT calculations are implemented by incorporating SOC, all the bands split into two bands. This effect is particularly strong in the conduction bands[30] as the conduction bands are derived predominantly from the heavy Pb atom that exhibits a large SOC effect.

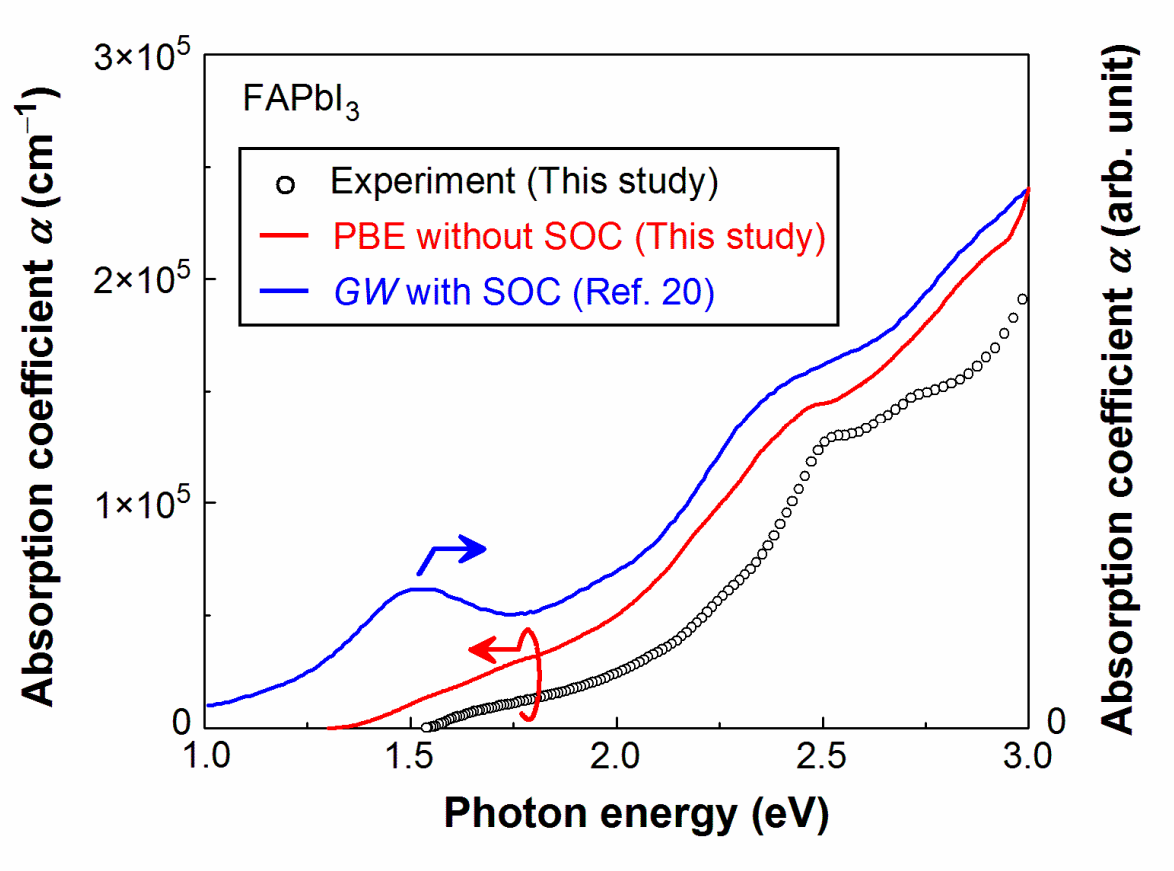

FIG. 10. $\alpha$ spectra of $FAPbI_3$ obtained from the experiment (open circles) and the DFT calculations based on the PBE without SOC (this study) and the *GW* method with SOC[20] (solid lines). The $\alpha$ spectrum of the PBE without SOC is obtained from $\varepsilon_c$ shown in Fig. 8(b).

Accordingly, the $\alpha$ peak observed at 1.5 eV in the *GW*-SOC result could be attributed to the effect of the conduction band splitting by SOC. It has been confirmed that the SOC-induced band splitting near the fundamental gap varies rather significantly with the orientation of the A-site cation[31,32,64] and the absorption characteristics near $E_g$ may be influenced strongly by the position and orientation of the A-site cation.

## C. Effects of A and X on the light absorption

The quite strong anisotropy observed in the DFT-derived dielectric functions in Fig. 8 can be understood by considering the strong interaction between the A-site cation and the surrounding X-site halogen atoms. Figure 11 shows the charge density profiles of (a) $\alpha$-$FAPbI_3$ and (b) $MAPbI_3$ in the energy region of −0.6 ~ −0.4 eV from VBM ($E$ = 0 eV). In Fig. 11, the charge density decreases with the order of red > yellow > green > blue with red being the highest. The charge density profiles in Fig. 11 represent the valence charge densities for the $E_2$ transition in $\alpha$-$FAPbI_3$ and the $E_1$ transition in $MAPbI_3$ at the $M$ point ($V_1C_1$). More specifically, in the case of $\alpha$-$FAPbI_3$, the energy position of $V_1$ at the $M_3$ point is −0.52 eV from VBM [Fig. 9(a)] and thus the valence charge density in this energy regime is responsible for the $E_2$ transition that occurs at 2.34 eV [see Fig. 9(c)]. For the $E_1$ transition of $MAPbI_3$ at 2.46 eV, the position of $V_1$ at the $M_2$ point is located at a similar energy of −0.44 eV from VBM.[5] Here, we particularly consider the charge density distribution at the $M$ point, as the visible light absorption in the hybrid perovskites is determined essentially by the interband transitions near the $M$ point ($V_1C_1$). The result of $MAPbI_3$ in Fig. 11(b) is obtained from the optimized $MAPbI_3$ structure reported in our previous study.[5]

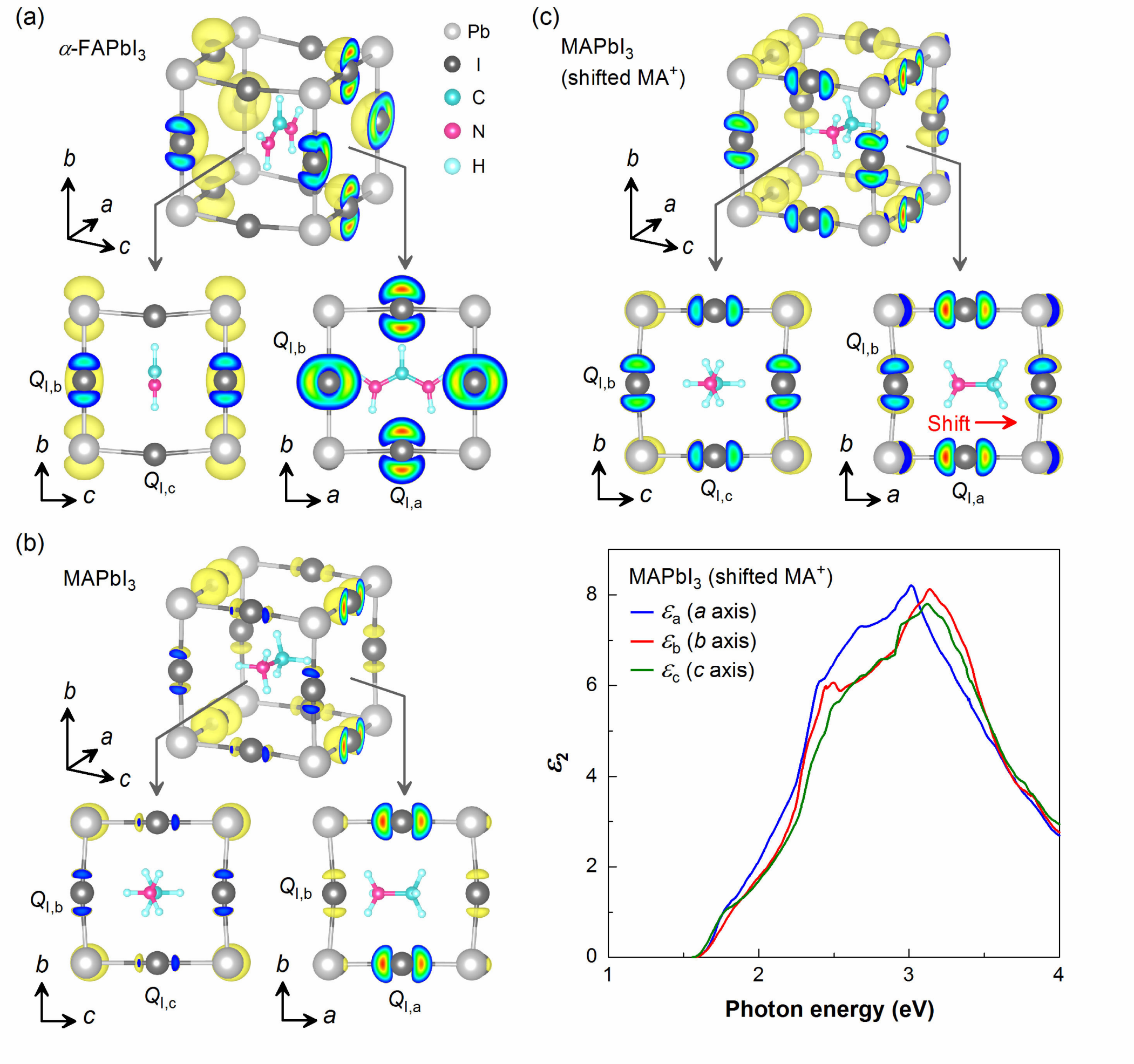

FIG. 11. Charge density profiles of (a) $\alpha$-$FAPbI_3$, (b) $MAPbI_3$ and (c) hypothetical $MAPbI_3$ with shifted $MA^+$ in the energy region of –0.6 ~ –0.4 eV from VBM ($E$ = 0 eV). These profiles represent the valence charge densities for the $E_2$ transition ($\alpha$-$FAPbI_3$) and the $E_1$ transition ($MAPbI_3$) at the $M$ point ($V_1C_1$). The charge density decreases with the order of red > yellow > green > blue with red being the highest. The valence charges of the I atoms along the $a$, $b$, and $c$ axes are indicated as $Q_{I,a}$, $Q_{I,b}$ and $Q_{I,c}$, respectively. In (c), the N atom of $MA^+$ is shifted to the center position of the C–N bond in (b) and the $\varepsilon_2$ spectra obtained from this hypothetical $MAPbI_3$ structure are also shown.

In Figs. 11(a) and 11(b), the charge density profiles in the same energy region (–0.6 ~

−0.4 eV) are shown, but the distributions of the I-5*p* charge density in $\alpha$-$FAPbI_3$ and $MAPbI_3$ are significantly different. Specifically, the I-5*p* valence electron of $\alpha$-$FAPbI_3$ concentrates on the I atoms along the *a* and *b* axes, and the valence charges of these I atoms ($Q_{I,a}$ and $Q_{I,b}$, respectively) are far larger than that of the *c* axis direction ($Q_{I,c}$). In contrast, the valence charge distribution is more uniform in $MAPbI_3$, although $Q_{I,a}$ is larger than $Q_{I,b}$ and $Q_{I,c}$.

We find that the distribution of $Q_{I,a\text{-}c}$ is determined by the strong interaction between the N and I atoms. It is now established that, in $MAPbI_3$ (Refs. 20, 32, 35-39) and $FAPbI_3$ (Refs. 20, 50), hydrogen bonding is formed between the I and H–N, which can be expressed as I($\delta^-$)···H($\delta^+$)–N($\delta^-$). Here, $\delta^+$ and $\delta^-$ represent the positive and negative partial charges, respectively. The absolute value of $\delta$ is governed predominantly by the electronegativity, and the hydrogen bonding of I···H–N leads to the reduction of the N-I distance, as confirmed from Fig. 11(b). Our systematic DFT calculations reveal that the distribution of $Q_{I,a\text{-}c}$ is controlled primarily by the N-I distance and $Q_{I,a\text{-}c}$ reduces significantly when the N-I distance is smaller (anti-coupling effect). In $\alpha$-$FAPbI_3$, for example, $Q_{I,c}$ becomes quite small due to the intense anti-coupling effect induced by the shorter N-I distance, and the valence charge concentrates on the *a-b* plane. In $MAPbI_3$, larger $Q_{I,a}$, compared with $Q_{I,b}$ and $Q_{I,c}$, can also be interpreted by the strong anti-coupling interaction as the N atom locates close to the *b-c* plane due to the effect of hydrogen bonding.

The highly anisotropic optical transitions at the *M* point can be explained almost completely by the anti-coupling effect. For the $E_2$ transition of $\alpha$-$FAPbI_3$ at 2.34 eV, for example, $\varepsilon_b$ shows the highest $\varepsilon_2$ value, while $\varepsilon_c$ indicates the lowest value [Fig. 8(b)]. As confirmed from Fig. 11(a), $Q_{I,a}$ shows the highest value among $Q_{I,a\text{-}c}$ and the charge density distribution of $Q_{I,a}$ is oriented along the *b* axis, which in turn increases the polarizability in this direction significantly. Consequently, $\varepsilon_b$ of $\alpha$-$FAPbI_3$ exhibits the maximum $\varepsilon_2$ value, while very low $Q_{I,c}$ leads to the drastic reduction of the transition probability and results in quite small $\varepsilon_c$. In the case of $MAPbI_3$, the charge density profile of $Q_{I,a}$ is aligned along the *a* axis [Fig. 11(b)], therefore increasing the $\varepsilon_a$ value at 2.46 eV ($E_1$ transition), whereas we observe $\varepsilon_b \sim \varepsilon_c$ since $Q_{I,b} \sim Q_{I,c}$, as confirmed from Fig. 8(c).

To justify quite strong anti-coupling effect in the hybrid perovskites, the charge density profile of $MAPbI_3$ is calculated by changing the position of $MA^+$ [Fig. 11(c)]. For this calculation, the N atom position of $MA^+$ is shifted intentionally and is placed at the center position of the C–N bond in Fig. 11(b). In this hypothetical configuration, the charge density distribution of −0.6 ~ −0.4 eV becomes more uniform, compared with the case of Fig. 11(b), as $Q_{I,b}$ and $Q_{I,c}$ increase due to the increased N-I distance. More importantly, as the result of the charge redistribution, the strong anisotropic behavior disappears in the calculated dielectric functions [Fig. 11(c)]. This effect can be seen clearly by comparing the polarization-dependent $\varepsilon_2$ spectra of $MAPbI_3$ in Fig. 8(c) with those in Fig. 11(c).

One very important result in Fig. 8 is that, among the $\varepsilon_2$ spectra calculated for different light polarizations, the $\varepsilon_2$ spectrum having the lowest $\varepsilon_2$ values in the visible region shows the best match with the experimental data. This result indicates that the visible-light absorption in the actual hybrid perovskites is modified significantly by the anti-coupling effect. In other words, although the Pb-derived conduction band ($C_1$) is independent of A, the electronic states of the I and Br are influenced strongly by the hydrogen bonding of the A-site cation.

To clarify the anti-coupling interaction further, we calculate the dielectric functions of various $APbX_3$ perovskites with different A-site cations ($FA^+$, $MA^+$, $NH_4^+$, $Cs^+$) and X-site halogen atoms (I, Br, Cl). In particular, the $APbX_3$ dielectric functions are obtained by replacing the A-site cation in identical $PbI_3^-$, $PbBr_3^-$ and $PbCl_3^-$ structures to determine the effect of the A-site cation on the optical absorption. For the $PbX_3^-$ structures, we employ the configurations that are determined from the structural optimization of $MAPbX_3$. In our calculations, the position and orientation of the A-site cation within a fixed $PbX_3^-$ cage are optimized and, in the resulting $CsPbX_3$ structures, $Cs^+$ locates near the center position of the unit cell. In the calculation of $\alpha$-$FAPbX_3$, however, the $FA^+$ orientation in Fig. 8(a) is assumed because a quite different $FA^+$ orientation is obtained when $FA^+$ is relaxed within the $PbX_3^-$ derived from $MAPbX_3$.

Figure 12 shows the $\varepsilon_2$ spectra of (a) $\alpha$-$FAPbX_3$, (b) $MAPbX_3$ and (c) $CsPbX_3$ calculated from the above procedure, together with corresponding $\varepsilon_2$ spectra of (d) $\alpha$-$FAPbX_3$, (e) $MAPbX_3$ and (f) $CsPbX_3$ obtained by applying the sum rule, which is explained below. In Fig. 12, only the $\varepsilon_2$ spectra of $\varepsilon_c$ ($\alpha$-$FAPbX_3$) and $\varepsilon_b$ ($MAPbX_3$ and $CsPbX_3$) are shown and the bars indicate the energy positions of the $E_1$ transition at the $M_2$ point ($V_1C_1$). The calculated dielectric functions of $NH_4PbX_3$ are almost the same with those of $MAPbX_3$ shown in Fig. 12(b). As confirmed from Figs. 12(a)-(c), the overall $\varepsilon_2$-spectral shapes are similar when the $\varepsilon_2$ spectra of the same X are compared. However, the $\varepsilon_2$ amplitude varies rather significantly with the A-site cation and the $\varepsilon_2$ value for the $E_1$ transition, indicated by the bar in Fig. 12, increases from $\varepsilon_2 = 3.1$ ($\alpha$-$FAPbI_3$) to 4.9 ($CsPbI_3$). Although the dielectric functions of $CsPbX_3$ have not been reported, the shape of the absorbance spectrum reported for $CsPbI_3$ (Refs. 83, 84) is quite similar to that shown in Fig. 12(c).

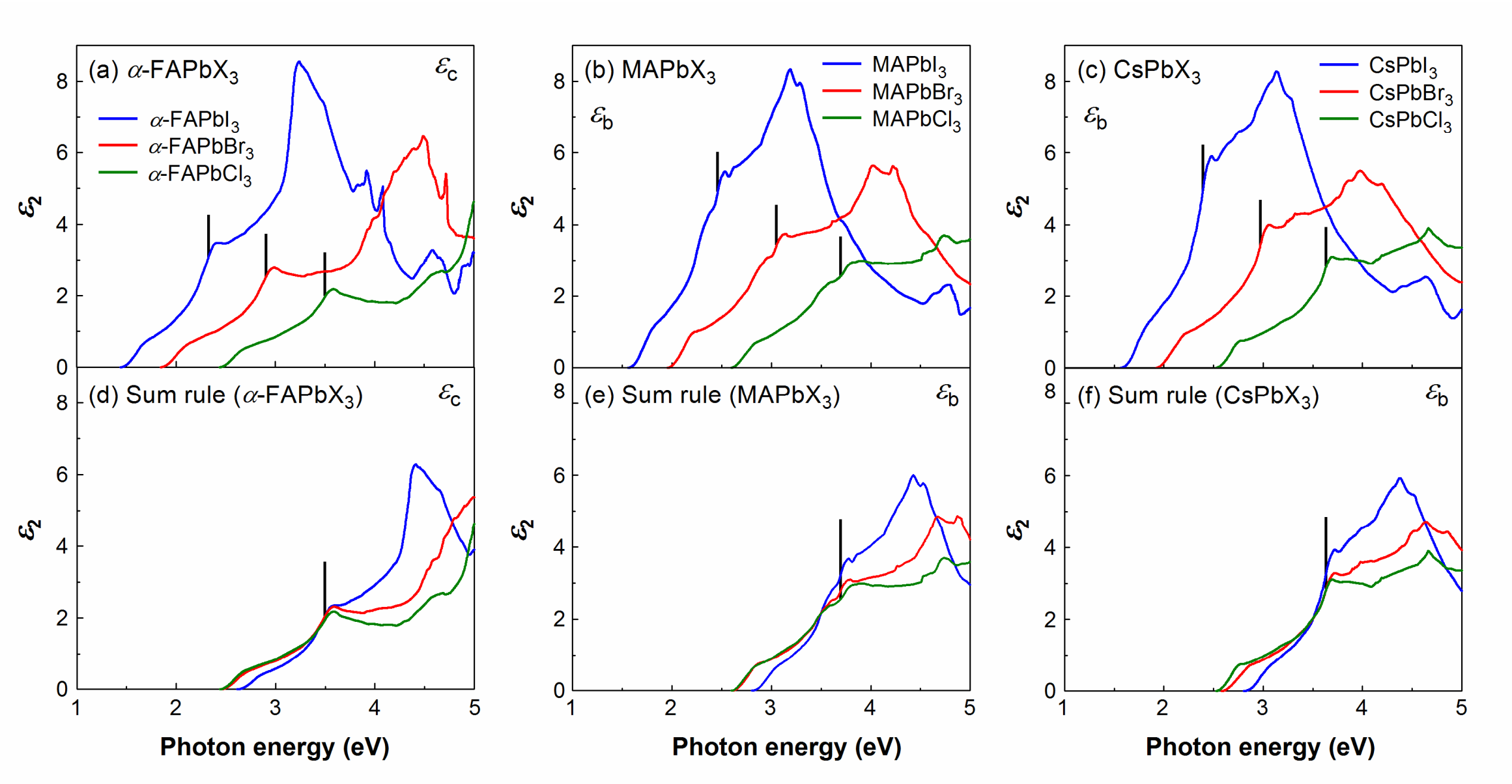

FIG. 12. $\varepsilon_2$ spectra of (a) $\alpha$-$FAPbX_3$, (b) $MAPbX_3$, and (c) $CsPbX_3$ (X = I, Br, Cl) calculated from the DFT, together with the corresponding $\varepsilon_2$ spectra of (d) $\alpha$-$FAPbX_3$, (e) $MAPbX_3$, and (f) $CsPbX_3$ obtained by applying the sum rule for $APbI_3$ and $APbBr_3$. In the figure, only the $\varepsilon_2$ spectra for $\varepsilon_c$ ($\alpha$-$FAPbX_3$) and $\varepsilon_b$ ($MAPbX_3$ and $CsPbX_3$) are shown. The bars indicate the energy positions of the $E_1$ transition at the $M_2$ point ($V_1C_1$). The dielectric functions of (a)-(c) have been obtained by replacing the A-site cation in the identical $PbX_3^-$ structures determined from the structural optimization of $MAPbX_3$.

To express the variation of the $\varepsilon_2$ spectrum with the A-site cation, we extract the $\varepsilon_2$ value for the $E_1$ transition at the $M_2$ point: i.e., the $\varepsilon_2$ value indicated by the bar in Figs. 12(a)-(c), for example. In Fig. 13(a), the extracted $\varepsilon_2$ values of $APbX_3$ are summarized as a function of the valence charge ratio expressed by $Q_X/Q_{Total}$. Here, $Q_X$ shows the valence charge of the X atom along the $a$, $b$, and $c$ axes in a selected energy region ($Q_{I,a\text{-}c}$ in Fig. 11, for example), while $Q_{Total}$ denotes the total charge of $APbX_3$ in the same energy region. Thus, $Q_X/Q_{Total}$ represents the relative magnitude of each X-site valence charge (i.e., $Q_{X,a}$, $Q_{X,b}$ or $Q_{X,c}$). For the $Q_X/Q_{Total}$ calculation, we select the energy ranges of –0.6 ~ –0.4 eV ($APbI_3$), –0.7 ~ –0.5 eV ($APbBr_3$) and –0.8 ~ –0.6 eV ($APbCl_3$) from VBM ($E$ = 0 eV), which correspond to the $E_1$ transitions at the $M_2$ point. It should be noted that, when the $PbX_3^-$ structures derived from $MAPbX_3$ are assumed, the energies of the $E_1$ and $E_2$ transitions in $\alpha$-$FAPbX_3$ become almost identical since the lattice parameters are quite similar in $MAPbX_3$. In Fig. 13(a), the results for $\varepsilon_a$, $\varepsilon_b$ and $\varepsilon_c$ in each $APbX_3$ are plotted.

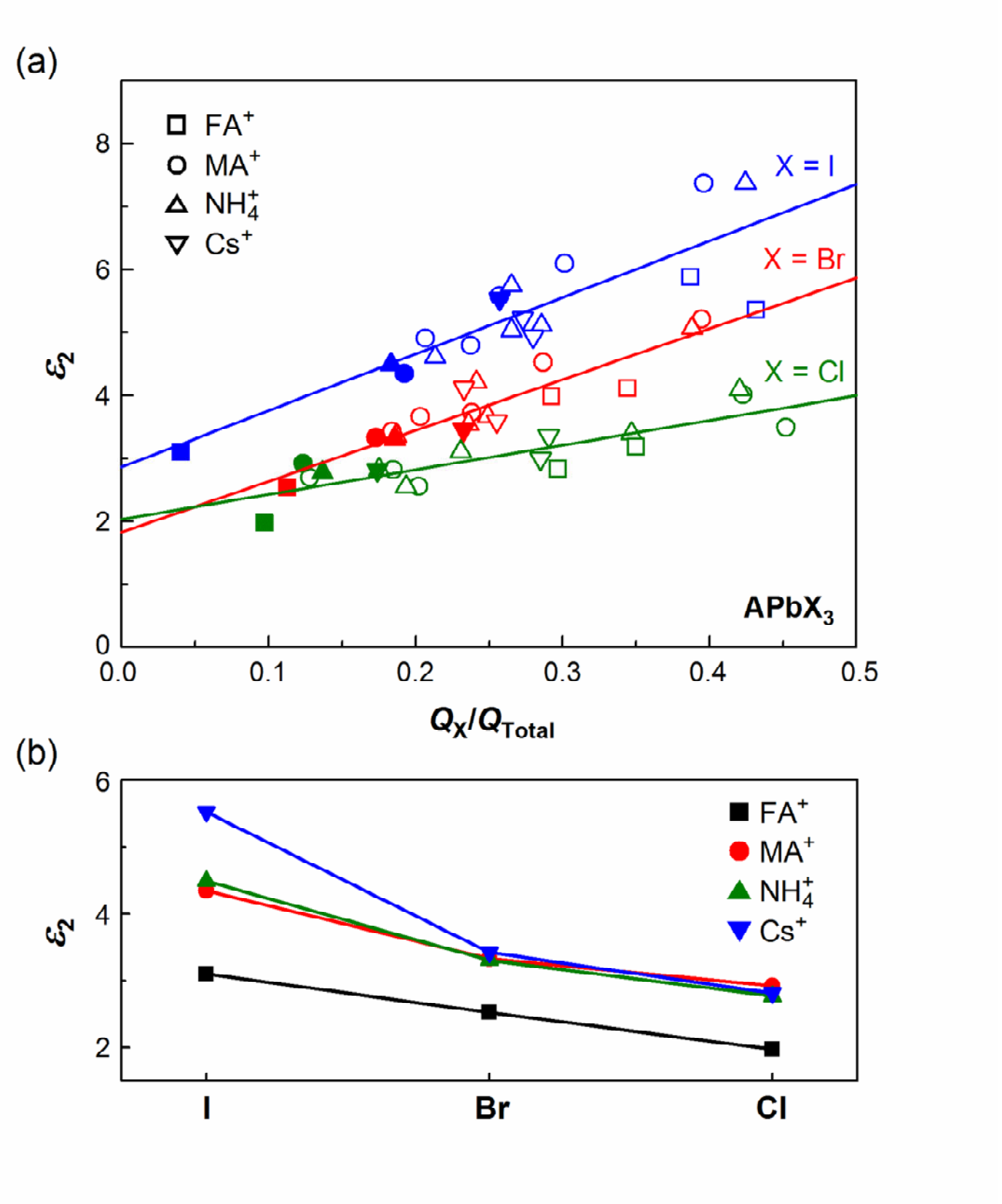

FIG. 13. (a) $\varepsilon_2$ values for the $E_1$ transition at the $M_2$ point ($V_1C_1$) in $APbX_3$ (A = $FA^+$, $MA^+$, $NH_4^+$, $Cs^+$; X = I, Br, Cl) as a function of the valence charge ratio expressed by $Q_X/Q_{Total}$ and (b) $\varepsilon_2$ value for the lowest $Q_X/Q_{Total}$ in each $APbX_3$. The $Q_X$ represents the valence charge of the X-site atom along the *a*, *b*, and *c* axes and is calculated by selecting an energy region that corresponds to the $E_1$ transition at the $M_2$ point ($V_1C_1$), whereas $Q_{Total}$ denotes the total charge of $APbX_3$ in the same energy region. In (a), the closed symbol represents the plot for the lowest $Q_X/Q_{Total}$ in each $APbX_3$. In (b), the $\varepsilon_2$ values of the closed symbols in (a) are summarized.

For $MAPbX_3$ and $NH_4PbX_3$ in Fig. 13(a), the DFT calculations are also implemented by varying the N atom position within the $PbX_3^-$. Specifically, in this calculation, the N atom of $MA^+$ is shifted to the center position of the C–N bond in the optimized $MAPbX_3$ structure [see Fig. 11(c)]. In $NH_4PbX_3$, the N atom position of $NH_4^+$ is quite similar to that of $MA^+$,[5] and we also calculate the dielectric function of $NH_4PbX_3$ when $NH_4^+$ is placed at the center position of the unit cell.

In Fig. 13(a), the linear increase of $\varepsilon_2$ with $Q_X/Q_{Total}$ is evident and the $\varepsilon_2$ value of $\varepsilon_a$, $\varepsilon_b$ or $\varepsilon_c$ in $APbX_3$ increases when the corresponding $Q_{X,a}$, $Q_{X,b}$ or $Q_{X,c}$ becomes larger.

The systematic variations of $\varepsilon_2$ with $Q_X/Q_{Total}$ are represented quite well by the liner fitting results (solid lines), confirming the presence of a universal rule for the effect of the A-site cation. In Fig. 13(a), the slopes for the $\varepsilon_2$ change with $Q_X/Q_{Total}$ are quite similar in $APbI_3$ and $APbBr_3$, while the $\varepsilon_2$ variation is suppressed in $APbCl_3$. The scatter of the plots reflects the fact that $Q_X$ varies slightly by the X–Pb–X bond angle and bond length.

The closed symbol in Fig. 13(a) represents the $\varepsilon_2$ value for the lowest $Q_X/Q_{Total}$ in each $APbX_3$ and these values are summarized in Fig. 13(b). Since the experimental optical properties are reproduced when the strong anti-coupling effect is considered, the $\varepsilon_2$ values of Fig. 13(b) approximate the light absorption strength in the hybrid perovskites. It is obvious that $\alpha$-$FAPbX_3$ shows the lowest $\varepsilon_2$, compared with the other $APbX_3$ perovskites. Because $FA^+$ has two N atoms, a quite large anti-coupling interaction occurs in $\alpha$-$FAPbX_3$, which in turn reduces the visible-light absorption significantly, as observed in Fig. 4. In contrast, $Cs^+$ is stabilized at the center position of the $PbX_3^-$ cage and shows the weak anti-coupling effect. As a result, $CsPbI_3$ exhibits the highest $\varepsilon_2$ value among all the $APbX_3$ perovskites investigated here. This result is consistent with the observed increase in $\alpha$ by the incorporation of Cs into $\alpha$-$FAPbI_3$.[53]

We further find that the effect of X on the $\varepsilon_2$ spectrum can be expressed simply by the general sum rule:[70]

$$\int E\varepsilon_2(E)dE = const. \qquad (1)$$

As shown in Figs. 4 and 12(a)-(c), the replacement of X induces the spectrum shift and the whole $\varepsilon_2$ spectrum moves toward higher energy with decreasing halogen mass. This change can be reproduced quite well based on the sum rule. In particular, the sum rule requires that the $\varepsilon_2$ peak amplitude reduces to $E/(E + \Delta E)$ when the $\varepsilon_2$ spectrum is shifted toward higher energy by $\Delta E$. In Figs. 12(d)-(f), the $\varepsilon_2$ spectra of $APbI_3$ and $APbBr_3$ are shifted so that the $E_1$ transition energies match that of $APbCl_3$. In this case, the amplitude of the shifted $\varepsilon_2$ spectra is reduced further by applying the sum rule. Under these assumptions, the $\varepsilon_2$ values for the $E_1$ transition become similar and all the spectra show reasonable overlap in the energy region of $E < 3.7$ eV. Accordingly, the $\varepsilon_2$ amplitude variation induced by X can simply be interpreted by the change in the transition energy. From Fig. 13(b), it can be seen that the $\varepsilon_2$ values reduce systematically with the order of I > Br > Cl, independent of the species of the A-site cation.

The above results indicate that the optical effects of the A-site cation and X-site halogen atom can be described separately according to the valence charge interaction and the sum rule, respectively. As a result, the optical interaction of the A-site cation in $APbX_3$ perovskites can be categorized into two main factors: (i) slight $E_g$ variation controlled primarily by the size of the A-site cation[17] and (ii) large variation of $\alpha$ induced by the anti-coupling effect. In $\alpha$-$FAPbI_3$, for example, $E_g$ decreases slightly compared with $MAPbI_3$, but $\alpha$ decreases rather significantly by the strong anti-coupling effect. In contrast, the introduction of $Cs^+$ increases $\alpha$ because of the weaker anti-coupling effect with a slight increase in $E_g$.

## VI. DISCUSSION

In polycrystalline materials, optical anisotropy is generally not observed and the optical properties are expressed as the average response of each anisotropic component. In the case of the polycrystalline hybrid perovskites, therefore, the dielectric function corresponding to $(\varepsilon_a + \varepsilon_b + \varepsilon_c)/3$ is expected to be observed experimentally. Rather surprisingly, the experimental $\varepsilon_2$ spectra of $\alpha$-$FAPbI_3$, $MAPbI_3$ and $MAPbBr_3$ do not match those calculated assuming optical isotropy [i.e., $(\varepsilon_a + \varepsilon_b + \varepsilon_c)/3$] and show excellent agreement with the DFT spectra obtained by maximizing the anti-coupling effect. In other words, the overall visible absorption in these perovskites is minimized by the strong anti-coupling interaction.

In our previous study for $MAPbI_3$,[5] the large difference observed between the experimental and isotropic DFT $[(\varepsilon_a + \varepsilon_b + \varepsilon_c)/3]$ spectra was attributed to the rapid reorientation of the A-site cation. It is now well established that $MA^+$ and $FA^+$ in the $PbI_3^-$ reorient rapidly with a time scale of 0.5−14 ps at room temperature.[33-35,40-42,62] In the DFT calculation that assumes 0 K temperature, however, the cation position is fixed completely and the effect of the cation reorientation is neglected. Accordingly, the excellent agreement between the experimental and DFT results supports the fact that the N-I distance is close and the anti-coupling phenomenon persists even at room temperature even though the A-site cation reorients rapidly.

This puzzling result can be understood from the quite strong hydrogen-bonding interaction between the I and H–N. In particular, the molecular dynamics (MD) simulations of $MAPbI_3$ show that the orientation of the center cation is not completely random even at room temperature[33-35,41,42,68,69] and the C-N axis of $MA^+$ is preferentially directed toward the face center position of the cubic structure.[33,41,42,68,69] In fact, the results obtained from the neutron scattering[41] and ultrafast infrared-vibrational spectroscopy[42] of $MAPbI_3$ are consistent with the MD simulation results, supporting the preferential orientations of $MA^+$ within the $PbI_3^-$ cage. The MD simulation of $MAPbI_3$ further reveals that the stable I-H distance of 2.65 Å is maintained at room temperature (268 K) due to the strong coupling of $MA^+$ with $PbI_3^-$.[34] Moreover, a recent large-scale MD simulation of $MAPbI_3$ shows that roughly half of $MA^+$ are participating in hydrogen bonding.[35] The above results show that dynamical motion of $MA^+$ is strongly correlated with the I atoms even at elevated temperatures due to the presence of I···H–N hydrogen bonding. Thus, if the close I-N distance is maintained statistically even during the continuous reorientation, the electrostatic anti-coupling interaction is still expected to occur.

A recent MD study further shows that the N-(C)-N axis of $FA^+$ is directed toward the center of the cube face [see Fig. 8(a)], and $FA^+$ rotates preferentially around the N-(C)-N axis.[62] This simulation confirms quite intense hydrogen-bonding interaction in $\alpha$-$FAPbI_3$ and suggests that the free motion of $FA^+$ is hindered effectively by the I···H–N bonding that occurs at both ends of $FA^+$. Accordingly, the significant anti-coupling effect observed in the room-temperature optical spectrum of $\alpha$-$FAPbI_3$ can be interpreted by the strong hydrogen-bonding effect of $FA^+$.

## VII. CONCLUSION

To determine the optical effect of the A-site cation in $APbI_3$-type hybrid perovskites, the reliable optical constants of $\alpha$-$FAPbI_3$ are determined without exposing the samples to humid air based on a self-consistent SE analysis. We find that the replacement of A = $MA^+$ with $FA^+$ reduces the overall amplitude of the $\varepsilon_2$ spectrum roughly by half and the resulting $\alpha$ values of $\alpha$-$FAPbI_3$ become notably smaller than those of $MAPbI_3$. The CP analysis shows that $E_g$ of $\alpha$-$FAPbI_3$ is 1.55 ± 0.01 eV, and $\alpha$-$FAPbI_3$ exhibits the sharp absorption edge with a low Urbach energy of 16 meV.

From the DFT calculations within the PBE without SOC, the optical transitions of $\alpha$-$FAPbI_3$ and $MAPbBr_3$ are analyzed. Although the DFT-derived dielectric functions indicate highly anisotropic optical properties, the calculated dielectric functions show excellent agreement with the experimental spectra, reproducing all the fine-absorption features observed in $\alpha$-$FAPbI_3$ and $MAPbBr_3$. For the assignment of the optical transitions, the dielectric response of each interband transition is calculated and the interband transitions at high symmetry points have been determined. The DFT calculation of the dielectric function further reveals that non-excitonic light absorption occurs in $\alpha$-$FAPbI_3$.

Our analysis for the $APbI_3$ perovskites shows that the strong optical anisotropy confirmed in the DFT results originates from the charge density distribution of the I-5*p* valence electron. In particular, the DFT calculation indicates that the formation of the hydrogen bonding between the I and H–N modifies the electronic state of the I atoms significantly and the valence charge density of the I atom reduces when the I-N distance is smaller. This anti-coupling effect is found to govern the visible light absorption and the transition probability reduces significantly when the anti-coupling interaction becomes stronger. Based on this finding, we establish a general rule that expresses the optical effect of the A-site cation (A = $FA^+$, $MA^+$, $NH_4^+$, $Cs^+$) within the $PbX_3^-$ cage (X = I, Br, Cl). Since $FA^+$ has two N atoms, the anti-coupling interaction is maximized, lowering $\alpha$ values in $\alpha$-$FAPbI_3$, while $CsPbI_3$ shows the highest $\alpha$ due to the weak anti-coupling effect of $Cs^+$. The systematic DFT calculations further show that the variation of the $APbX_3$ dielectric function with X can be approximated by the sum rule. From the above results, we propose universal rules that describe the effects of A and X on the visible light absorption separately. Our results demonstrate that experimental dielectric functions of various organic-inorganic hybrid perovskites can be predicted based on the polarization-dependent DFT calculations.